\documentclass{article}

\usepackage{amsmath,amsfonts,amsbsy,amssymb,fullpage,makeidx,array,graphicx}
\usepackage{microtype}
\usepackage{amsmath}
\usepackage{amssymb}
\usepackage{mathtools}
\usepackage{rotating}
\usepackage{epstopdf}
\usepackage{caption}
\usepackage{lipsum}
\usepackage{afterpage}
\usepackage{bm}
\usepackage{subcaption}
\usepackage{pstricks}
\usepackage{pdftricks}

\usepackage{ulem} 

\def \eqref#1{(\ref{#1})}
\def \meanw{\langle w \rangle}
\begin{document}

\begin{flushleft}
{\Large
\textbf{One-Dimensional Population Density Approaches to Recurrently Coupled Networks of Neurons with Noise}
}

\vspace{0.5cm}

{\large
Wilten Nicola$^{1,\ast}$, 
Cheng Ly$^{2,\ast}$, 
Sue Ann Campbell$^{1,\ast}$\\
}
\vspace{0.5cm}

{\small
\bf{1} Department of Applied Mathematics, University of Waterloo, Waterloo, Ontario Canada
\\
\bf{2} Department of Statistical Sciences and Operations Research, Virginia Commonwealth University, Richmond, Virginia 23284-3083  U.S.A.
\\
$\ast$ E-mail: wnicola@uwaterloo.ca, cly@vcu.edu, sacampbell@uwaterloo.ca
}

\end{flushleft}

\begin{center}
\today
\end{center}

\section*{Abstract}
Mean-field systems have been previously derived for networks of coupled, two-dimensional, integrate-and-fire neurons such as the Izhikevich, adapting exponential (AdEx) and quartic integrate and fire (QIF), 
among others.  Unfortunately, the mean-field systems have a degree of frequency error and the networks analyzed often do not include noise when there is adaptation.  
Here, we derive a one-dimensional partial differential equation (PDE) 
approximation for the marginal voltage density under a first order moment closure for coupled networks of integrate-and-fire neurons with white noise inputs.  
The PDE has substantially less frequency error than the mean-field system, and provides a great deal more information, at the cost of analytical tractability.  The convergence properties of the mean-field system 
in the low noise limit are elucidated.  A novel method for the analysis of the stability of the asynchronous tonic firing solution is also presented and implemented.  Unlike previous attempts at stability analysis 
with these network types, information about the marginal densities of the adaptation variables is used.  This method can in principle be applied to other systems with nonlinear partial differential equations.

\section{Introduction}

The population density approach is a commonly used framework for analyzing large networks of model neurons \cite{AbbottandVreeswijk,at06,HM1,HM,Knight2,ly_tranchina_07,us2}.  Rather than 
tracking the individual behavior of neurons, a probability density function (PDF) for each population is considered.  The PDF represents the probability that 
any individual neuron is in a particular state, or, equivalently, the proportion of neurons in the population that have the particular state.  The population density equation usually takes the form of a partial differential equation for the probability density of the voltage and other neuronal variables.   Unfortunately, the population density 
equation has as many dimensions as the individual neuronal equations, and often has complicated boundary conditions.  Thus, the more complex the neural model, the more difficult it is to both analyze 
and solve the associated population density equation.

Fortunately, a great deal of the rich dynamics displayed by real neurons can be replicated via suitably complex, two-dimensional integrate-and-fire models.  This class of models includes the Izhikevich model 
\cite{Izhikevich}, the Adaptive Exponential model (AdEx) \cite{AdEx}, and the Quartic model \cite{Touboul2008}, to name a few.  These models represent an excellent trade off in the sense that they are simple, 
discontinuous oscillators, however once properly fit, they can predict the spike times and membrane potential of actual neurons with a great deal of accuracy.  

The population density equations generated by networks of these neurons are still exceptionally difficult to analyze and numerically simulate.  Thus mean-field equations for these types of networks were derived \cite{us2}.  The derivation uses a sequence of analytical reductions, including a first order moment closure, and a separation of time scales to obtain a small system of ordinary differential equations for certain moments of the network from the original population density equation.   The mean-field system is usually very accurate for slow behaviors, such as bursting oscillations, and for steady state and transient firing dynamics.  Additionally, being a simple set of ordinary differential equations, the mean-field equations are easily solved using any standard integration scheme.
However, due to the approximations made in the process of the derivation, the mean-field system cannot provide detailed information about fast-oscillations or network level synchrony.  There is also a marked error in the frequency of bursting observed in the mean-field system compared to full network simulations \cite{us2}. 

Given the overly complex two-dimensional population density equations, and the inability of the mean-field system to capture the full dynamics of the networks, 
here we suggest a reduction of the full population density equation to
a one-dimensional PDE coupled to a system of ODEs.
The PDE-ODE system, which is derived assuming first order moment closure, is simpler to solve numerically and to analyze. In addition, it drastically minimizes the bursting frequency error present in the mean-field system \cite{us2}, and is able to predict rapid behaviors while providing information about the synchrony of the network.  In particular, we find that this system
robustly captures the complex temporal dynamics exhibited in simulations of the networks. 
While moment-closure methods have been analyzed in \cite{ly_tranchina_07}, the networks were primarily leaky integrate-and-fire networks.  This paper considers several neuron 
models with two dimensions: voltage and adaptation. The neurons all receive 
external white noise forcing and are in all-to-all coupled networks.  
The bulk of the numerical simulations and results have been obtained with the Izhikevich model \cite{Izhikevich}.  For the purposes of comparison however, the general mean-field system which applies to any of the models is derived.  

In section \ref{models}, we introduce the class of networks we are considering, followed by their respective population density equations in section \ref{pde}.  The first order moment closure approximation is applied in section \ref{momclos} to derive the coupled PDE-ODE system. A closed form expression for the 
steady state solution of the PDE is found and used with a quasi-steady state approximation derived the associated mean-field model.  In this section we will also present various results about the boundary conditions used in the population density equations and the convergence 
of the mean-field system  in the low noise limit. 
Numerical simulation examples for several model types are presented in section \ref{numerics}, while a novel 
stability analysis method is presented in section \ref{stability} that qualitatively captures some of the features of the network.  This method can potentially be applied to other systems with nonlinear partial differential 
equations of two state variables.

\subsection{2-Dimensional Neural Models with White Noise}\label{models}

\begin{figure}
        \centering
        \begin{subfigure}[b]{0.85\textwidth}
                \centering
                \includegraphics[width=\textwidth]{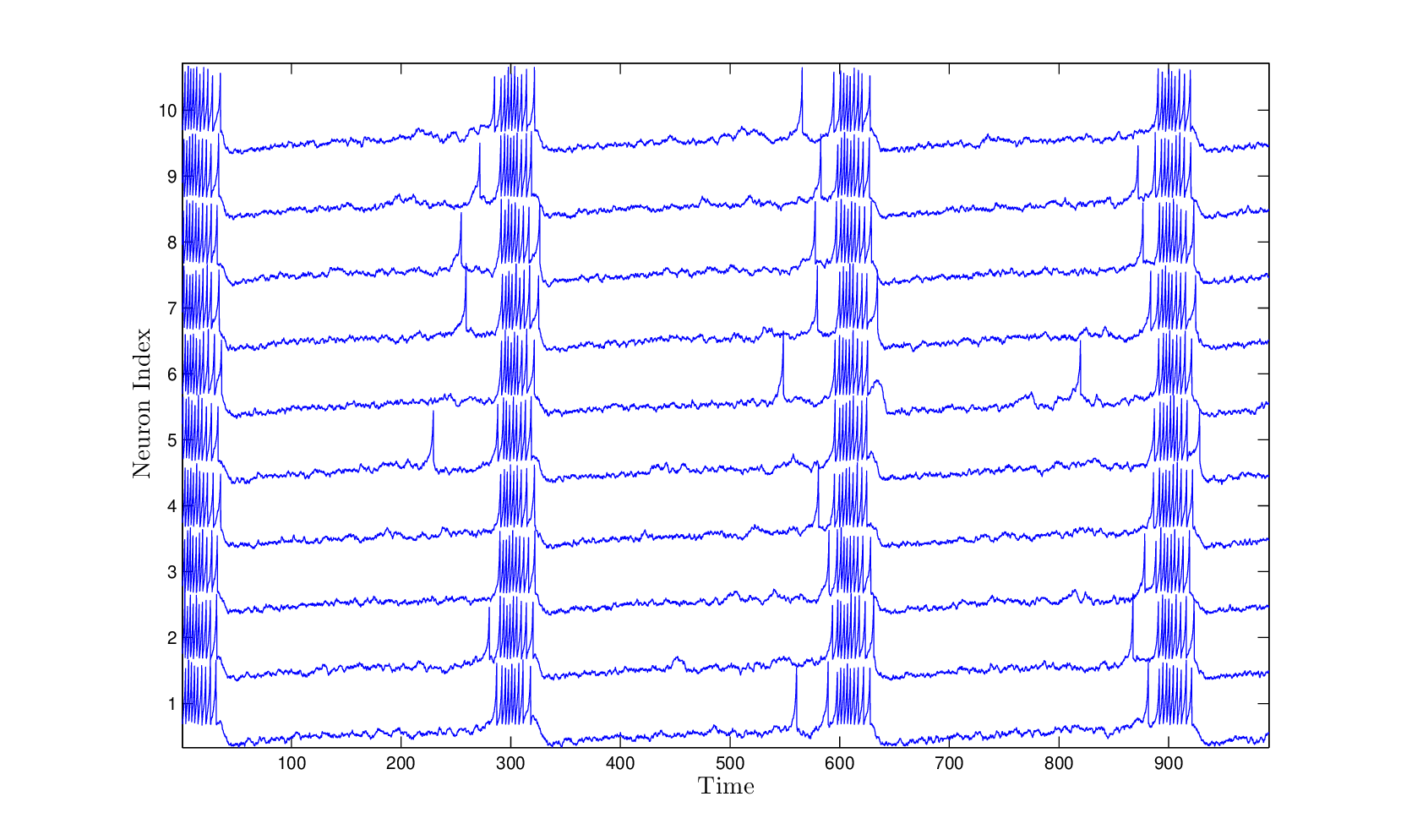}
                \caption{$v(t)$ for 10 randomly selected neurons }   
                \label{FIGTP1}
        \end{subfigure}
\quad 
        \begin{subfigure}[b]{0.85\textwidth}
                \centering
                \includegraphics[width=\textwidth]{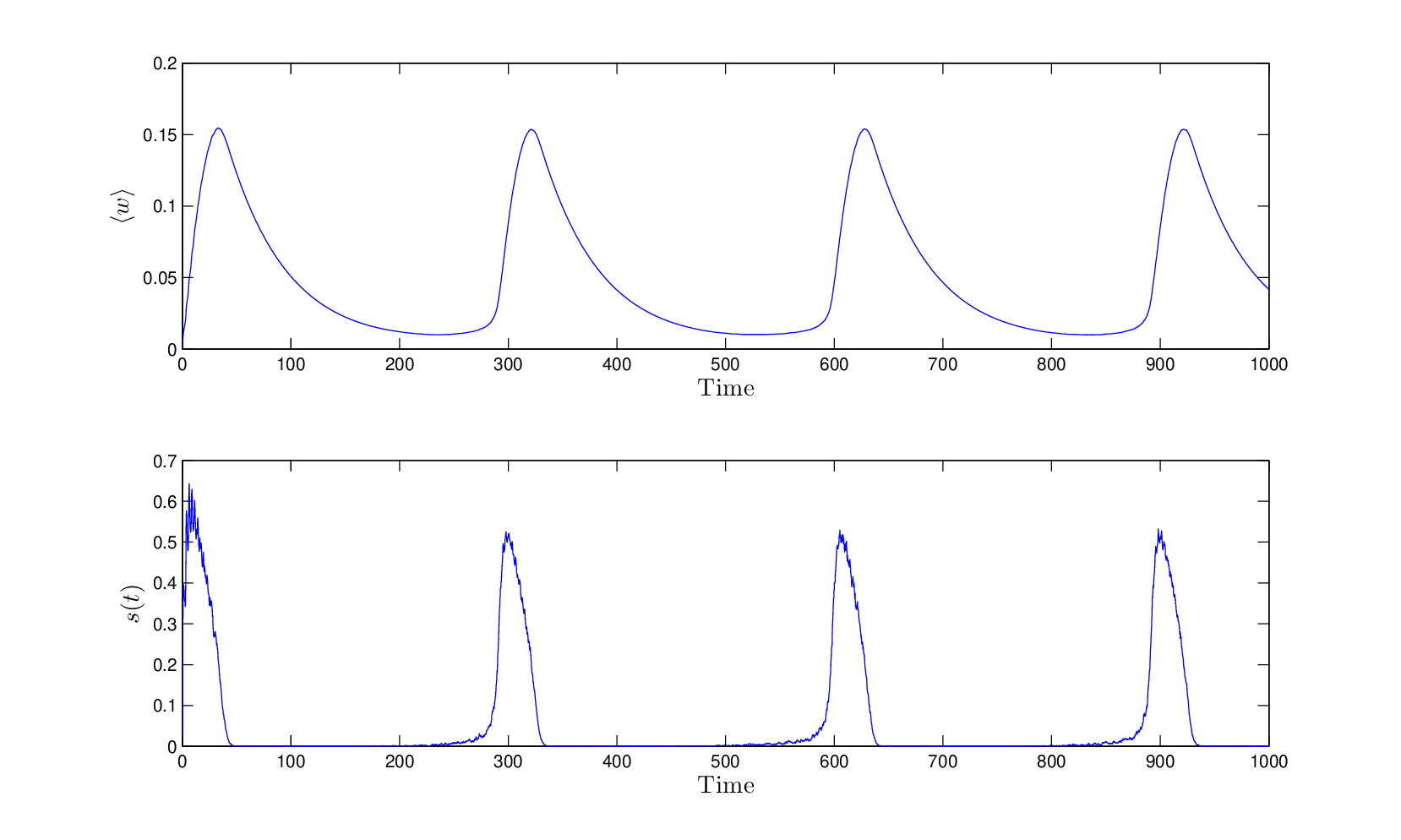}
                \caption{$\langle w \rangle$ and $s(t)$}
                \label{FIGTP2}
        \end{subfigure}
\caption{Numerical simulation of a network of 1000 all-to-all excitatory 
coupled Izhikevich neurons with white noise currents using the intrinsically bursting parameter set (IB) in Table \ref{table_param}.  Other parameter values
are $g=1.111$, $I=0.035$, and $\sigma=0.04$.  (a) Voltage traces for 10 
randomly selected neurons.  (b) Network average adaptation and synaptic activity. The synchronous bursting is induced by the noise as the mean-current level is below rheobase.  }  \label{FIGTP}
\end{figure}

The set of models we consider are all-to-all recurrently coupled networks described by the following equations: 
\begin{eqnarray}
\dot{v}_i &=& F(v_i) - w_i + I + gs(e_r-v_i) +\eta_i = G_v(v_i,s,w_i) + \eta_i \label{mod1}\\
\dot{w}_i &=& \frac{W_\infty(v_i) - w}{\tau_W(v_i)} =  G_w(v_i,w_i) \label{mod2} \\
\dot{s} &=& -\frac{s}{\tau_s} + \frac{s_{jump}}{N} \sum_{j=1}^N \sum_{t<t_{j,k}}\delta(t-t_{j,k}) \label{mod3}
\end{eqnarray}
where $v_i$ is the scaled dimensionless voltage, $w_i$ is a recovery/adaptation variable (for $i=1,2,\ldots N$), $s$ is population averaged synapse variable, and $t_{j,k}$ is the $k$th spike fired by the $j$th neuron in the network.  The quantity $\eta_i$ is a gaussian white noise process that models the large amount of random inputs neurons receive, with 
$$ \langle \eta_i(t) \rangle = 0 ,\qquad \langle \eta_i(t_p)\eta(t_q) \rangle = \sigma^2 \delta(t_p-t_q).$$ 
Additionally, the variables $v$ and $w$ have the following resets/jumps: 
\begin{equation} v(t^-) = v_{peak} \Rightarrow \begin{cases} v(t^+) = v_{reset} \\ w(t^+) = w(t^-) + w_{jump} \end{cases} \end{equation}
This is a fairly broad class of models, that includes various subtypes, such as: 
\begin{eqnarray*}
F(v) &=& -\frac{v}{\tau_m}, \quad \text{(Leaky Integrate and Fire)} ,\\
F(v) &=& v(v-\alpha), \quad \tau_W(v) = \tau_w, \quad W_\infty(v) = bv , \quad \text{(Izhikevich)}, \\
F(v) &=& e^v - v, \quad  \tau_W(v) = \tau_w, \quad W_\infty(v) = bv, \quad \text{(Adaptive Exponential)}, \\
F(v) &=& v^4 - \frac{2v}{\tau_w}, \quad \tau_W(v) = \tau_w, \quad  W_\infty(v) = bv, \quad \text{(Quartic Integrate and Fire)} \\
\end{eqnarray*}
Additionally, the Izhikevich model has various modifications aside from the default form presented above, all of which fall under the general set of equations given by (\ref{mod1})-(\ref{mod3}).   While the majority of these models are relatively new, they are readily being fit to neural data recordings, and to describe a wide variety of network level phenomena \cite{AdEx,us,Katie,Izhikevich}. 
For example, networks of these neurons burst for a large variety of parameter sets, as shown in Figure \ref{FIGTP}, and in \cite{us,nesse08,us2}.  Despite the abundance of noise in neural networks, the aforementioned models are often analyzed without noise.  Some exceptions include \cite{nesse08} which considered a leaky integrate-and-fire 
network with adaptation and slow noise, and \cite{richardson09} which considered an adaptive exponential integrate-and-fire model with noisy voltage.  

\subsection{Population Density Methods}\label{pde}

For networks with a large number of neurons ($N\rightarrow \infty$), the behavior of the population can be described by a probability density function, $\rho(v,w,t)$, where 
\begin{eqnarray*}
\int_\Omega \rho(v,w,t)\,dvdw = P( (v_i(t),w_i(t)) \in \Omega)
\end{eqnarray*}
i.e., integration of the probability density function over a subset $\Omega$ of state space gives the probability a neuron in the network is in the region $\Omega$.  
In the large 
network limit, one can rigorously derive a population density equation for the network of neurons. 
The evolution equation for $\rho(v,w,t)$ is:
\begin{eqnarray}
\frac{\partial \rho(v,w,t)}{\partial t} = -\nabla \cdot \bm J(v,w,s,t)\label{pde1}
\end{eqnarray}
where
\begin{eqnarray}
\bm J(v,w,s,t) &=& \begin{pmatrix}J^V(v,w,s,t) \\ J^W(v,w,t)\end{pmatrix} \\
J^V(v,w,s,t) &=&G_v(v,s,w)\rho(v,w,t) - \frac{\sigma^2}{2}\frac{\partial \rho(v,w,t)}{\partial v} \label{fj}\\
J^W(v,w,t)&=& G_w(v,w)\rho(v,w,t)\label{fw}.
\end{eqnarray}
Additionally, the discontinuities in the integrate-and-fire models result in boundary conditions on the probability flux:
\begin{eqnarray}
J^V(v_{peak},w,s,t)&=& \lim_{v\rightarrow v_{reset}^+}J^V(v,w+w_{jump},s,t)- \lim_{v\rightarrow v_{reset}^-}J^V(v,w+w_{jump},s,t) \label{bc1}\\
J^W|_{\partial W} &=& 0 \label{bc2} 
\end{eqnarray}
This yields a discontinuous flux term, due to the reset. Note that if we force $v\in[v_{reset},v_{peak}]$ by implementing a boundary on the neurons when $v = v_{reset}$  in addition to the typical reset at $v=v_{peak}$, then we can simply rewrite the boundary condition as 
\begin{equation} 
J^V(v_{peak},w,s,t) = J^V(v_{reset},w+w_{jump},s,t),\label{truebc1}
\end{equation}
 as done in \cite{us2}.  Further, numerical simulation of the population density 
equation requires a restriction in the domain which we choose to be $[v_{reset},v_{peak}]$ for convenience. Thus in the rest of the paper, we will assume  
$v\in[v_{reset},v_{peak}]$. For the sake of completeness, however, we include a derivation of the mean-field system on the unrestricted domain $-\infty<v\le v_{peak}$ in Appendix A.  

In the large network limit, one can also show that $s(t)$ converges to the ODE: 
\begin{eqnarray}\label{syn_eqn}
\dot{s} = -\frac{s}{\tau_s} + s_{jump} \int_W J^V(v_{peak},w,s,t)\,dw 
\end{eqnarray}
where the integral term is the network averaged firing rate $\nu(t)$ \cite{us2}.   
To summarize, we have the following PDE/ODE coupled system: 
\begin{eqnarray}
\frac{\partial \rho(v,w,t)}{\partial t} &=& -\frac{\partial}{\partial v} \left( (F(v) - w + gs(e_r-v)+I)\rho(v,w,t) - \frac{\sigma^2}{2}\frac{\partial \rho(v,w,t)}{\partial v}\right) \label{pdet1}\\ &-&\frac{\partial}{\partial w}\left(\left(\frac{ W_\infty(v)-w}{\tau_W(v)}\right)\rho(v,w,t)\right) \\
\dot{s} &=& -\frac{s}{\tau_s} + s_{jump} \int_W J^V(v_{peak},w,s,t)\,dw \label{odet1}
\end{eqnarray}
subject to the boundary conditions (\ref{bc1})-(\ref{bc2}). This system is fairly difficult to solve beyond some rudimentary first order methods.   However, there are analytical techniques that substantially reduce the complexity of the PDE.  The technique that we will employ is a moment closure \cite{ly_tranchina_07}.  The general principle of such dimension reduction methods have been applied to the 
statistics of network connectivity \cite{liu_nykamp_09, rangan09} and to master equations of stochastic networks \cite{bressloff09,buice10}.

\section{First and Higher Order Moment Closure}\label{momclos}
The population density equation is equivalent to the marginal voltage density multiplied by the conditional $w$ density: 
\begin{eqnarray} \rho(v,w,t) = \rho_W(w|v,t)\rho_V(v,t) \label{cond1}\end{eqnarray}
Substituting this into eq. (\ref{pde1}), integrating with respect to $w$ and using the boundary condition (\ref{bc2}), we arrive at the one-dimensional PDE: 
\begin{eqnarray}
\frac{\partial \rho_V(v,t)}{\partial t} &=& -\frac{\partial}{\partial v}\left[\rho_V(v,t)\left(F(v)-\langle w|v\rangle + I + gs(e_r-v)\right) - \frac{\sigma^2}{2}\frac{\partial \rho_V(v,t)}{\partial v}\right]\label{1dpde}\\
&:=& - \frac{\partial J(v,\langle w | v\rangle,s,t)}{\partial v} 
\end{eqnarray}
where the flux, $J$ has been redefined and $\langle w | v\rangle$ is the conditional mean of $w$ given $v$.   Additionally, the equation for $s$ becomes 
\begin{eqnarray}
\dot{s} &=& -\frac{s}{\tau_s} + s_{jump} J(v_{peak},\langle w | v_{peak}\rangle,s,t) \nonumber \\
     &=& -\frac{s}{\tau_s} + s_{jump} \left( (F(v_{peak}) - \langle w| v_{peak}\rangle + gs(e_r-v_{peak}) + I)\rho_V(v_{peak},t) - \frac{\sigma^2}{2}\left.\frac{\partial \rho_V(v,t)}{\partial v}\right|_{v_{peak}} \right) \label{newsde}.
\end{eqnarray}
Integration with respect to $w$ is also needed to derive a new boundary condition on $\rho_V(v,t)$. Starting with the right-hand side of (\ref{truebc1}):
\begin{eqnarray*}\label{bc_1stMomClos} 
\int_W J^V(v_{reset},w+w_{jump},s,t)\,dw &=& \int_W G_V(v_{reset},s,w+w_{jump})\rho_W(w+w_{jump}|v_{reset},t)\rho_V(v_{reset},t)\,dw \\&-& \frac{\sigma^2}{2}\int_W \rho_V(v_{reset},t)\frac{\partial \rho_W(w+w_{jump}|v,t)}{\partial v}\bigg|_{v_{reset}}\,dw  \\&-& \frac{\sigma^2}{2}\int_W\rho_W(w+w_{jump}|v_{reset},t) \frac{\partial \rho_V(v,t)}{\partial v}\bigg|_{v_{reset}} \,dw.
\end{eqnarray*}
Note that $\rho_W(w+w_{jump}|v)$ is merely the conditional density in $w$ shifted by $w_{jump}$ to the left.  Thus, since we are still integrating over the entire $w$ domain, we have the following: 
\begin{eqnarray*} 
 \int_W \rho_W(w+w_{jump}|v_{reset},t) \rho_V(v_{reset},t)G(v_{reset},s,w+w_{jump})\,dw &=& G(v_{reset},s,\langle w | v_{reset}\rangle) \rho_V(v_{reset},t)\\
\int_W \rho_V(v_{reset},t)\frac{\partial \rho_W(w+w_{jump})}{\partial v}\bigg|_{v=v_{reset}}\,dw&=& \rho_V(v_{reset},t)\frac{\partial}{\partial v}\left(\int_W  \rho_W(w+w_{jump})\,dw \right) \bigg|_{v_{reset}} =0\\
\int_W\rho_W(w+w_{jump},t|v_{reset}) \frac{\partial \rho_V(v,t)}{\partial v}\bigg|_{v=v_{reset}} \,dw &=& \frac{\partial \rho_V(v,t)}{\partial v}\bigg|_{v_{reset}}
\end{eqnarray*} 
It follows that 
$$ \int_W J^V(v_{reset},w+w_{jump},s,t) = G(v_{reset},s,\langle w| v_{reset}\rangle)\rho_V(v_{reset},t) -\frac{\sigma^2}{2}\frac{\partial \rho_V(v,t)}{\partial v} \bigg|_{v_{reset}} = J(v_{reset},s,\langle w | v_{reset}\rangle,t) $$
Similar integration steps show that 
$$ \int_W J^V(v_{peak},w,s,t) = G(v_{peak},s,\langle w| v_{peak}\rangle)\rho_V(v_{peak},t) -\frac{\sigma^2}{2}\frac{\partial \rho_V(v,t)}{\partial v} \bigg|_{v_{peak}} = J(v_{peak},s,\langle w | v_{peak}\rangle,t)  $$
and the boundary condition becomes
\begin{equation}
 J(v_{reset},s,\langle w | v_{reset}\rangle,t) = J(v_{peak},s,\langle w|v_{peak}\rangle,t)
\end{equation}

So far every step applied has been exact and no approximation has been made.  However, without a PDE for $\langle w | v\rangle$, one cannot solve the PDE \eqref{1dpde} for $\rho_V$.   Using the probability density function (\ref{pde1}), one can derive a PDE for the quantity  $\langle w | v\rangle \rho_V(v,t) \left(=\int w\rho(v,w,t)\,dw\right)$:
\begin{eqnarray}
\frac{\partial}{\partial t}\left(\rho_V(v,t)\langle w| v\rangle \right) &=& -\frac{\partial}{\partial v}\left[\langle w|v\rangle \left( F(v)+gs(e_r-v)+I\right)\rho_V(v,t) - \langle w ^2 |v\rangle \rho_V(v,t) - \frac{\sigma^2}{2}\frac{\partial \langle w|v \rangle \rho_V}{\partial v} \right] \nonumber \\&-& \left(\frac{\langle w | v\rangle - W_\infty(v)}{\tau_W(v)} \right)\rho_V(v,t)\label{pdemom1}
\end{eqnarray}
There are two issues with this equation.  The first is that we would need to divide by $\rho(v,t)$ to isolate for $\langle w | v\rangle$, which yields problems when $\rho(v,t) = 0$ \cite{ly_tranchina_07}.  The second is that the presence of $\langle w ^2 |v\rangle$ necessitates yet another 1-dimensional PDE for $\langle w^2 | v\rangle\rho_V(v,t)$ and in general, the PDE of the $n^{\text{th}}$ conditional moment contains the $n+1^{\text{st}}$ conditional moment.  An approximation is necessary to end the dependence of the $\langle w^n|v\rangle$ moment on 
$\langle w ^{n+1}|v\rangle$, i.e., to close the system.  Moment closure approximations in general assume a relationship between the higher moments with the lower moments.  We will consider two cases, the noiseless network ($\sigma =0$) and the network with noise $\sigma>0$.  

\subsection{The $\sigma = 0$ Case} 

For the $\sigma = 0$ case, one can apply a higher order moment closure assumption with a straightforward physical meaning.  In particular, making the assumption 
\begin{eqnarray}
\langle w^2|v \rangle - \langle w|v\rangle^2 = \sigma^2_{w|v} = 0 \label{HOMC}
\end{eqnarray}
we have the following: 
\begin{eqnarray}
\frac{\partial}{\partial t}\left(\rho_V(v,t)\langle w| v\rangle \right) &=& -\frac{\partial}{\partial v}\left[\langle w|v\rangle G_v(v,s,\langle w | v\rangle) \rho_V(v,t) \right] +   \left(\frac{\langle w | v\rangle - W_\infty(v)}{\tau_W(v)} \right)\rho_V(v,t) \nonumber \\
&=& \frac{\partial \rho_V(v,t)}{\partial t}\langle w | v\rangle - G_v(v,s,\langle w|v\rangle)\rho_V(v,t)\frac{\partial \langle w |v\rangle}{\partial v}  +   \left(\frac{\langle w | v\rangle - W_\infty(v)}{\tau_W(v)} \right)\rho_V(v,t) \nonumber\\
\rho_V(v,t)\frac{\partial \langle w | v\rangle}{\partial t}&=&- \rho_V(v,t)G_v(v,s,\langle w|v\rangle)\frac{\partial \langle w |v\rangle}{\partial v}  -   \left(\frac{\langle w | v\rangle - W_\infty(v)}{\tau_W(v)} \right)\rho_V(v,t) \label{step23}
\end{eqnarray}
As every term in eq. \eqref{step23} contains $\rho_V(v,t)$, we can factor it out (assuming it is non zero on $[v_{reset},v_{peak}]$ for all $t$) which results in the following closed form equation for $\langle w | v\rangle$:
\begin{eqnarray}
\frac{\partial \langle w | v\rangle}{\partial t} =  -G_v(v,s,\langle w | v \rangle) \frac{\partial \langle w | v\rangle}{\partial v}- \frac{\langle w | v\rangle - W_\infty(v)}{\tau_w(v)}
\end{eqnarray}
Given the moment closure assumption \eqref{HOMC}, in addition to the reset in the voltage and the jump in $w$ at each spike, it is clear that the following boundary condition should apply: 
\begin{eqnarray}
\langle w | v_{reset}\rangle = \langle w | v_{peak}\rangle + w_{jump}
\end{eqnarray}
(see Appendix \ref{err} for a derivation).  Coupling this partial differential equation to the PDE \eqref{1dpde} for $\rho_V(v,t)$ with $\sigma=0$
and the ODE \eqref{newsde} for $s$ gives the following system: 
\begin{eqnarray}
\frac{\partial \rho_V}{\partial t} &=& -\frac{\partial}{\partial v}\left(G_v(v,s,\langle w | v \rangle)\rho_V\right) \label{bx1} \\
\frac{\partial \langle w | v\rangle}{\partial t} &=&  -G_v(v,s,\langle w | v \rangle) \frac{\partial \langle w | v\rangle}{\partial v}- \frac{\langle w | v\rangle - W_\infty(v)}{\tau_w(v)}\\
\dot{s}&=&-\frac{s}{\tau_S} + s_{jump} J(v_{peak},\langle w | v_{peak}\rangle,s,t) = -\frac{s}{\tau_S} + s_{jump} G_v(v_{peak},s,\langle w | v_{peak} \rangle)\rho_V(v_{peak},t)\label{bx3}
\end{eqnarray}
where $v\in[v_{reset},v_{peak}].$

One can interpret the assumption $\sigma^2_{w|v}  =0$ statistically as the random variable $w$ is a function of the random variable $v$, $w = g(v) = \langle w | v\rangle$ in which case the density in $w$ will be determined by the standard change of variables formula: 
$$ \rho_W(w) = \rho_V(g^{-1}(w))\left|\frac{d}{dw}(g^{-1}(w))\right|.$$
We have simulated this system for the noiseless network, and it improves on the first order moment closure approach in the noiseless case by providing more details and accuracy of the distribution of $w$ by accurately approximating $\langle w|v\rangle$.  However, the situation is more complicated once noise is added to the network.  

\subsection{The $\sigma > 0$ Case}

Returning to eq. \eqref{pdemom1} with $\sigma>0$ and applying the moment closure 
assumption (\ref{HOMC}) results in the following simplified equation for 
$\rho_V(v,t)\langle w | v\rangle$: 
\begin{equation}\label{secondPDE_noise}
	\rho_V \frac{\partial \langle w | v \rangle}{\partial t} =-\rho_V G_v(v,s,\langle w | v\rangle) \frac{\partial \langle w | v\rangle}{\partial v}
	-\frac{\langle w | v\rangle - W_\infty(v)}{\tau_w(v)}\rho_V
	+\frac{\sigma^2}{2} \left( \rho_V \frac{\partial^2 \langle w | v\rangle}{\partial v^2} + \frac{\partial \rho_V}{\partial v}\frac{\partial \langle w | v\rangle}{\partial v} \right)
\end{equation}
Unfortunately, unlike the noiseless case, $\rho_V$ is not a factor in every term in particular it is not a factor of the last term in eq. \eqref{secondPDE_noise}, and thus it cannot be removed from the equation.  This results in a substantially  more complicated and possibly ill-posed system.   Thus, we will not employ this moment closure assumption when $\sigma >0$.  However, a potential avenue of future research is to use a perturbation approach to study the solutions of this PDE in the low noise limit.  

As an alternative, we will use a standard first order moment closure assumption, given by 
\begin{equation}
\langle w|v\rangle = \langle w \rangle, \label{HOMC2}
\end{equation}
which reduces the PDE for $\rho_V(v,t)$  to 
\begin{eqnarray}
\frac{\partial \rho_V(v,t)}{\partial t} &=& - \frac{\partial J(v,\langle w \rangle,s,t)}{\partial v}. \label{pdet}
\end{eqnarray}
All that remains is to derive a differential equation for $\langle w \rangle$.  In particular, one can show that 
\begin{eqnarray}
\langle w \rangle' &=& \left\langle \frac{W_\infty(v) - w}{\tau_W(v)}\right\rangle  +w_{jump} J(v_{peak},\langle w|v_{peak}\rangle,s,t) + O(w_{jump}^2) \\
& \approx& \left\langle \frac{W_\infty(v) - w}{\tau_W(v)}\right\rangle  +w_{jump} J(v_{peak},\langle w\rangle,s,t)\\
& \approx&  \frac{\langle W_\infty(v)\rangle - \langle w\rangle}{\langle \tau_W(v)\rangle}+w_{jump} J(v_{peak},\langle w\rangle,s,t)\label{wde} 
\end{eqnarray}
where any function of $v$, $g(v)$, can be averaged using $\rho_V(v,t)$.  Combining (\ref{wde}) with the PDE (\ref{pdet}) for $\rho_V(v,t)$ and the ODE (\ref{odet1}) 
for $s$, gives the following system: 
\begin{eqnarray}
\frac{\partial \rho_V(v,t)}{\partial t} &=& - \frac{\partial}{\partial v}\left( (F(v) - \langle w\rangle + gs(e_r-v)+I)\rho_V(v,t) - \frac{\sigma^2} {2}\frac{\partial \rho_V(v,t)}{\partial v} \right) = -\frac{\partial}{\partial v} J(v,\langle w \rangle,s,t) \label{sys1} \\
\dot{\langle w \rangle} &=&  \frac{\langle W_\infty(v)\rangle - \langle w\rangle}{\langle \tau_W(v)\rangle}  +w_{jump} J(v_{peak},\langle w \rangle,s,t) \label{sys2}\\
\dot{s} &=& -\frac{s}{\tau_s} + s_{jump}J(v_{peak},\langle w \rangle,s,t) \label{sys3}
\end{eqnarray}
As in the $\sigma=0$ case, first order moment closure can be used derive to the boundary condition 
for the PDE \eqref{sys1}: 
\begin{equation}
J(v_{peak},s,\langle w \rangle ,t) = J(v_{reset},s,\langle w \rangle,t). \label{bc} 
\end{equation}
We note that a similar PDE/ODE system for $\rho_V(v,t)$ and $\langle w \rangle$ was derived in \cite{moritz} for an excitatory/inhibitory network of AdEx neurons.  The coupling used in \cite{moritz} was different from the synaptic coupling function $s$ considered here.  We will consider the application of this model to some examples in section~\ref{numerics}.

\subsection{Steady state density, boundary conditions and mean-field equations} 
\label{mfsec}
The coupled system of one PDE and two ordinary differential equations derived above 
is one step removed from a mean-field approximation.  In particular, if the variables $\langle w \rangle$ and $s$ operate on a slow enough time scale, then one can apply a separation of time scales to solve the PDE for $\rho_V(v,t)$ at steady state and hence solve for the $t-$independent flux: $J(v,\langle w \rangle,s)$. This in turn can be used to derive a two dimensional ODE mean-field model.

Assuming $\meanw$ and $s$ are fixed parameters, the steady-state solution of the one-dimensional 
PDE \eqref{sys1}--\eqref{sys3} must satisfy the following ordinary differential equation:
\begin{equation}
 0 = -\frac{\partial}{\partial v} \left[(F(v) - \langle w\rangle + gs(e_r-v)+I)\rho_V(v) - \frac{\sigma^2} {2}\frac{\partial \rho_V(v)}{\partial v}\right] = -\frac{\partial J(v,\langle w \rangle,s)}{\partial v}.  \label{ss1}
\end{equation}
It is clear from this equation that the boundary condition \eqref{bc} is automatically satisfied 
at steady state as the solution for $J(v,\meanw,s)$ is independent of $v$. Thus alternate
boundary conditions will be needed.
One can solve this ODE on the interval $(-\infty,v_{peak})$ or add a reflecting boundary condition at $v_{reset}$ and restrict the solution to $[v_{reset},v_{peak}]$.   Using the interval $(-\infty,v_{peak}]$, one obtains a solution which is continuous everywhere, but not differentiable at $v=v_{reset}$.  The solution is smooth if we restrict it to the interval $[v_{reset},v_{peak}]$, as we do henceforth.  
To explicitly show the dependence of the density and the firing rate on the noise level, we will temporarily write $\rho_V(v) = \rho_V(v;\sigma)$ and $\nu = \nu(\sigma)$

When the system has no noise, one can easily solve for the steady-state density, $\rho_V(v;0)=\rho_0(v)$, 
using only a boundary condition relating the flux to the firing rate, $\nu(0)=\nu_0$:
\[ J(v_{peak},\meanw, s)=\nu_0. \]
The firing rate can then be determined using the normalization condition
\begin{equation}
\int_{v_{reset}}^{v_{peak}} \rho_0(v)\, dv = 1. \label{normcond}
\end{equation}
Doing this one obtains: 
\begin{eqnarray*}
\rho_0(v) &=& \begin{cases} \frac{\nu_0}{G_v(v,s,\langle w \rangle)}  & I- I^*(s,\langle w \rangle)>0\\ 
\delta(v-v_-(s,\langle w \rangle)) & I- I^*(s,\langle w\rangle) \le 0 \end{cases} \\
\nu_0 &=& \begin{cases} \left[\int_{v_{reset}}^{v_{peak}}\frac{dv}{G_v(v,s,\langle w \rangle)}\right]^{-1}  & I - I ^*(s,\langle w \rangle )>0 \\ 0&I -I^*(s,\langle w \rangle)\le 0 \end{cases}
\end{eqnarray*}
where $I-I^*(s,\langle w\rangle)$ is the switching manifold for the system and is given by: 
\begin{equation}
I - I^*(s,\langle w \rangle) = \min_{v\in[v_{reset},v_{peak}]}\left[ G_v(v,s,\langle w \rangle)\right]
\end{equation}
and $v_-(s,\langle w \rangle)$ is the asymptotically stable equilibrium point that exists when 
$I-I^*\le 0$ for the DE
$$\dot{v} = F(v) - \langle w \rangle + gs(e_r-v) + I = G_v(v,s,\langle w \rangle).$$ 
with $s$ and $\langle w \rangle$ treated as parameters.

The resulting mean-field system is 
\begin{eqnarray}
\dot{s}& = &-\frac{s}{\tau_s} + s_{jump}\nu_0(s,\langle w \rangle) \label{mf01}\\
\dot{\langle w \rangle} &= &  \frac{\langle W_\infty(v)\rangle - \langle w\rangle}{\langle \tau_W(v)\rangle}  +w_{jump}\nu_0(s,\langle w \rangle)\label{mf02}
\end{eqnarray}
Note that this is a non-smooth system of differential equations.   As we shall see, the mean-field system for noise is a qualitatively different class of system because it is 
a completely smooth system of ODE's.  However, we will show how these two systems are related to one another in the $\sigma\rightarrow 0$ limit.  

To solve for the steady state of the system with noise, an additional boundary condition is required
as eq.\eqref{ss1} is a second order ODE.  The typical boundary conditions applied are:
\begin{eqnarray}
J(v_{peak},\langle w \rangle,s) &=& \nu(\sigma) \quad \text{(Definition of firing rate)}\label{nubc}\\
\rho_V(v_{peak};\sigma)&=& 0 \quad \text{(Absorbing Boundary Condition)}\label{absbc}
\label{jumpbc}
\end{eqnarray}
These boundary conditions have been previously used in \cite{fourcaudbrunel} in their analysis of the leaky integrate-and-fire models with white noise, and in \cite{AbbottandVreeswijk}.   We note that in these two papers the justification for the absorbing boundary condition appears to be different.  In \cite{fourcaudbrunel}, the justification is that $\rho_V(v;\sigma)=0$ for $v>v_{peak}$ and thus for continuity and integrability reasons, the authors set $\rho(v_{peak};\sigma)=0$.  In \cite{AbbottandVreeswijk}, the authors state that $\rho(v_{peak};\sigma)=0$ as all the firing is due to noise, and thus the deterministic component of the flux should not contribute anything.


In the following, we will derive the solution in some detail. This will allow us to offer an
alternative justification for the boundary conditions and to investigate the limiting behaviour
of $\rho(v_{peak};\sigma)$ and $\nu(\sigma)$ as $\sigma\rightarrow 0$.
We restrict ourselves to the case where $I>I^*(s,\langle w \rangle)$.  The case when $I<I^*(s,\langle w \rangle)$ is more complicated, but can be dealt with using the same approach. Solving equation 
(\ref{ss1}) for $\rho_V(v;\sigma)$ and using the boundary condition \eqref{nubc} yields:  
\begin{eqnarray*}
\rho(v;\sigma) &=&-\frac{2\nu(\sigma)}{\sigma^2}\int_{v_{reset}}^v \exp\left(-\frac{2}{\sigma^2}(M(v')-M(v))\right)\,dv' + D\exp\left(\frac{2}{\sigma^2}M(v) \right)
\end{eqnarray*}
where $M(v)$ is an anti-derivative of $F(v) -\langle w \rangle +gs(e_r-v) + I = G_v(v,s,\langle w \rangle)$ and $D=\rho(v_{reset})\exp(\frac{2}{\sigma^2}M(v_{reset}))$.
Before proceeding further, we use Laplace's method for integrals \cite{BO} to shed some insight into the asymptotic behavior of $\rho(v;\sigma)$.  In particular, note the following asymptotic behaviors that are valid if $I>I^*(s,\langle w \rangle)$:
 \begin{eqnarray*}
\frac{2}{\sigma^2}\int_{v_{reset}}^v \exp\left(-\frac{2}{\sigma^2}(M(v')-M(v))\right)\,dv'  &\sim& \frac{\exp(\frac{2}{\sigma^2}(M(v)-M(v_{reset})))}{G_v(v,s,\langle w \rangle)}, \quad \sigma\rightarrow 0 \\
\frac{2}{\sigma^2}\int_{v}^{v_{peak}} \exp\left(-\frac{2}{\sigma^2}(M(v')-M(v))\right)\,dv'  &\sim& \frac{1}{G_v(v,s,\langle w \rangle)}, \quad \sigma\rightarrow 0 \\
\end{eqnarray*}
 This would seem to imply that if $\nu(\sigma)$ is convergent in the $\sigma\rightarrow 0$ limit, the density function contains a divergent term as if $G_v(v,s,\langle w \rangle) >0$, then $M(v)>M(v_{reset})$ and the first term diverges exponentially fast as $\sigma \rightarrow 0$.    Thus, to obtain a convergent density function, we need to remove the first term in the integral.  

Rewriting the density as: 
\begin{eqnarray*}
\rho(v;\sigma) &=&\frac{2\nu}{\sigma^2}  \int_{v}^{v_{peak}} \exp\left(-\frac{2}{\sigma^2}(M(v')-M(v))\right)\,dv' + \left[D-\frac{2\nu}{\sigma^2}\int_{v_{reset}}^{v_{peak}} \exp\left(-\frac{2}{\sigma^2}M(v')\right)\,dv'\right]\exp\left(\frac{2}{\sigma^2}M(v) \right)
\end{eqnarray*}
Since we are still free to specify a boundary condition, we may choose $D$ (and hence $\rho(v_{reset})$)
to eliminate the divergent term, yielding:
\begin{equation}
 \rho(v;\sigma) =\frac{2\nu(\sigma)}{\sigma^2}  \int_{v}^{v_{peak}} \exp\left(-\frac{2}{\sigma^2}(M(v')-M(v))\right)\,dv'   \label{rho1}
\end{equation}
Note that this choice of $D$ is equivalent to applying the boundary condition \eqref{absbc}. Thus, 
the boundary condition can be seen as a regularity condition requiring the density $\rho(v;\sigma)$ 
be well behaved in the small noise limit.

As in the noiseless case, applying the normalization condition on $\rho(v;\sigma)$ yields an 
expression for the firing rate:
\begin{equation}
\nu(\sigma) = \left(\frac{2}{\sigma^2}\int_{v_{reset}}^{v_{peak}} \int_{v'}^{v_{peak}}\exp\left(-\frac{2}{\sigma^2}(M(v',\langle w \rangle,s)-M(v,\langle w \rangle ,s))\right)dv'dv\right)^{-1}. \label{nu1}
\end{equation}
This leads to the following mean-field system for the network:
\begin{eqnarray}
\dot{s}& = &-\frac{s}{\tau_s} + s_{jump}\nu(\sigma,s,\langle w \rangle) \label{mf1}\\
\dot{\langle w \rangle} &= & \frac{\langle W_\infty(v)\rangle - \langle w\rangle}{\langle \tau_W(v)\rangle}  +w_{jump}\nu(\sigma,s,\langle w \rangle).\label{mf2}
\end{eqnarray}

\begin{figure}
        \centering
        \begin{subfigure}[b]{0.65\textwidth}
                \centering
                \includegraphics[width=\textwidth]{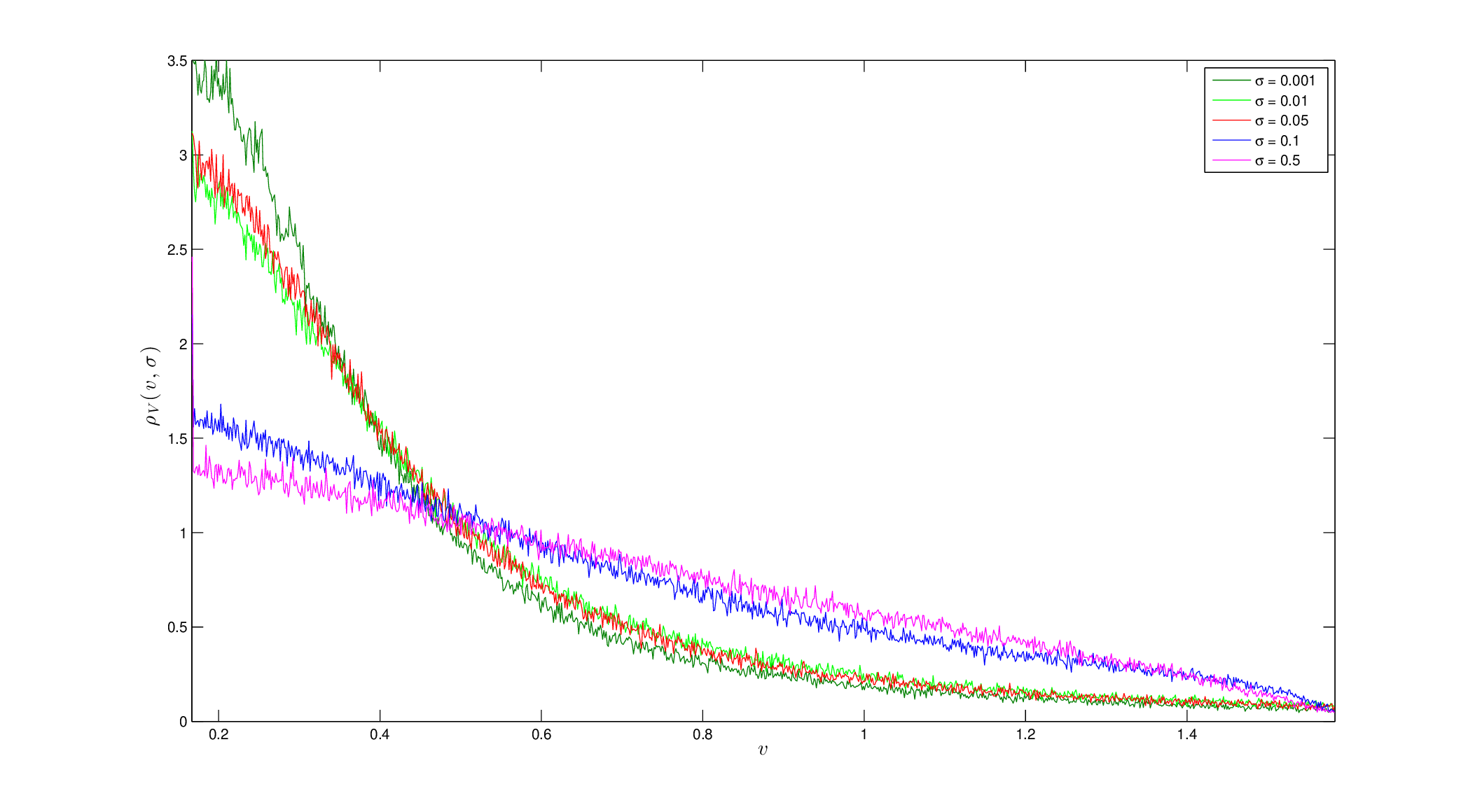}
                \caption{$\rho_V(v;\sigma)$ for various $\sigma$ from simulations of a network with 50,000 neurons}   
                \label{Density1}
        \end{subfigure}
\quad 
        \begin{subfigure}[b]{0.65\textwidth}
                \centering
                \includegraphics[width=\textwidth]{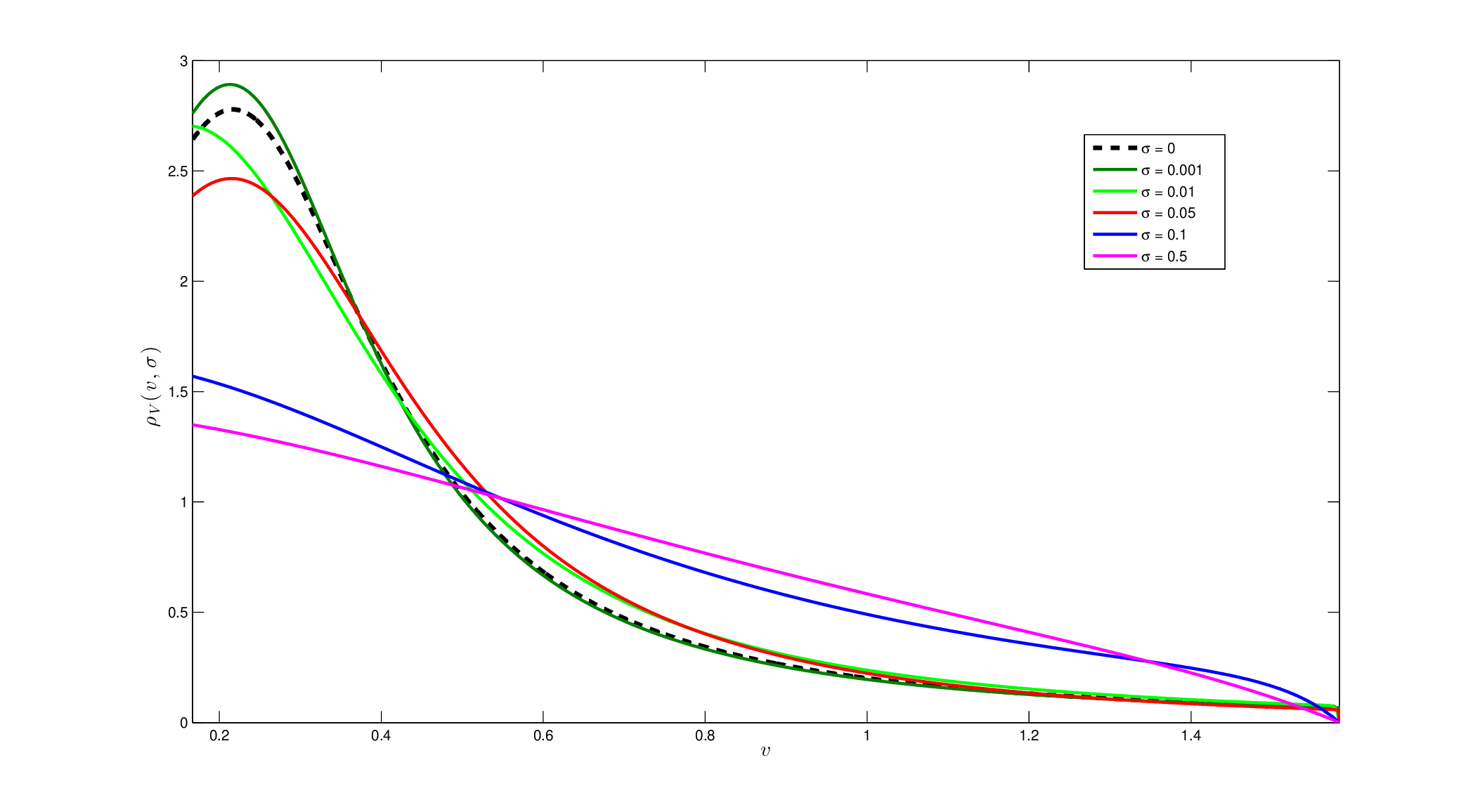}
                \caption{$\rho_V(v;\sigma)$ from eq. \eqref{rho1}}
                \label{Density2}
        \end{subfigure}\\

        \begin{subfigure}[b]{0.65\textwidth}
                \centering
                \includegraphics[width=\textwidth]{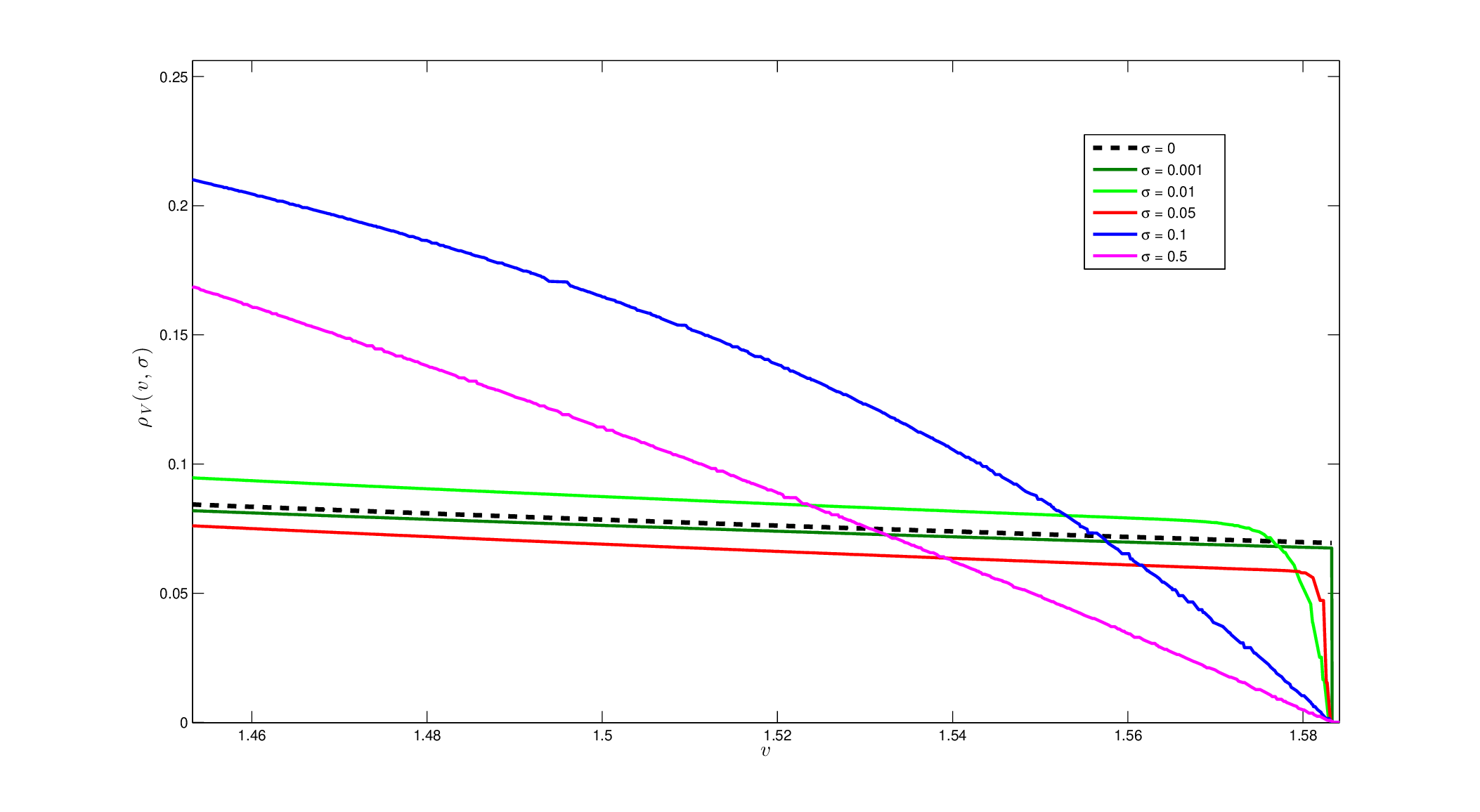}
                \caption{A close up of the convergence of the density $\rho_V(v;\sigma)$ as $\sigma\rightarrow 0$ near $v_{peak}$}  
                \label{Density3}
        \end{subfigure}
\caption{(a) A coupled network of 50,000 Izhikevich neurons was simulated until steady state and the steady state density $\rho_V(v;\sigma)$ was determined by using a normalized histogram.  (b) The solution for the steady-state density was found analytically using eq. \eqref{rho1}.  (c) The nature of the convergence 
of the density $\rho(v;\sigma)$ to $\rho_0(v)$, the analytical solution to the steady-state density without noise.  The density function $\rho(v;\sigma)$ only converges pointwise to $\rho_0(v)$ on $[v_{reset},v_{peak})$, with the derivative becoming unbounded at $v=v_{peak}$.  
 The parameters are the rapid spiking (RS) parameter set in Table \ref{table_param}., with $g,I$ chosen such that the steady state of the network was tonic firing. }  \label{FIGD}
\end{figure}

Using the expansions of the integrals given above shows that the solution has the 
following asymptotic behaviour
\begin{eqnarray*}
\rho(v;\sigma) &\sim& \rho_0(v)=\frac{\nu_0}{G_v(v,s,\langle w \rangle)} \quad \sigma \rightarrow 0\\ 
\nu(\sigma) &\sim& \nu_0 \quad \sigma\rightarrow 0
\end{eqnarray*}
for $I>I^*(s,\langle w \rangle)$. 
This implies that, in the tonic firing region of the parameter space, the
firing rate converges to the noiseless value, which in turn implies that the mean field
equations converge to the noiseless mean field model. The convergence of the density is more
delicate. Note that the firing rate and density at steady state are related by 
$$ \rho_0(v)G_v(v,s,\langle w \rangle) = \nu_0, \quad  I>I^*(s,\langle w \rangle)$$ 
for the noiseless network.  Since  $\nu_0>0$ when $I>I^*(s,\langle w \rangle)$,
the boundary condition \eqref{absbc} leads to an inconsistency at $v=v_{peak}$.
Thus the convergence of the density is only pointwise and for $v_{reset}\le v <v_{peak}$.
An example of this is shown in Figure \ref{FIGD}(c). 

This inconsistency can be dealt with by noting that $\rho(v_{peak};\sigma)=0$ is a sufficient, 
but not necessary condition for $\rho(v;\sigma)$ to converge to $\rho_0(v)$ for $v\neq v_{peak}$.  
In fact it can be weakened to yield convergence even at $v_{peak}$.  Specifically, making the 
following choice for $D$ 
\begin{eqnarray*}
D = \frac{2\nu}{\sigma^2}\int_{v_{reset}}^{v_{peak}} \exp\left(-\frac{2}{\sigma^2}M(v')\right)\,dv' + \exp\left(-\frac{2}{\sigma^2}M(v_{peak})\right)\rho_0(v_{peak})
\end{eqnarray*} 
one can show that the term 
\begin{eqnarray*}
\exp\left(\frac{2}{\sigma^2}(M(v) - M(v_{peak}))\right)\rho_0(v_{peak})
\end{eqnarray*} 
added to the density converges to $\rho_0(v_{peak})$ if $v=v_{peak}$ and 0 otherwise.  Thus, the criteria $\rho(v_{peak};\sigma)=0$ is not necessary even for convergence at 
$v=v_{peak}$ as $\sigma\to0$.  

The point here is not to use alternate solutions for the density and the firing rate, but rather to demonstrate that the absorbing boundary condition is sufficient and illustrate that the mean-field system
does converge to the mean-field system without noise for $I>I^*(s,\langle w \rangle)$.  A similar approach when $I<I^*(s,\langle w \rangle)$ demonstrates the same convergence.   Thus, solutions of the non-smooth noiseless mean-field system (\ref{mf01})-(\ref{mf02}) could be used as order zero solutions in a weak noise perturbation expansion of solutions of the mean-field system above.  We remark that the noiseless mean-field system has an analytically tractable bifurcation structure \cite{us3}.  We leave analysis of the bifurcation structure of the mean-field system with noise for future work.

\subsection{Numerical Examples} \label{numerics}
In this section, we compare simulations of the PDE system (\ref{sys1})--(\ref{sys3}), the mean-field system (\ref{mf1})-(\ref{mf2}) and of the full 
network \eqref{mod1}-\eqref{mod3} with 10,000 neurons.  

We begin by considering different parameter sets for the Izhikevich model, 
taken from \cite{IzBook}, which were fit to data for various neuron types.   
We use parameter sets for the CA1 pyramidal neuron, the intrinsically 
bursting neuron (IB), the chattering neuron (CH), and the rapidly spiking 
neuron (RS). The parameter values are given in Table~\ref{table_param}. As 
illustrated in 
Figure~\ref{FIGWV} for the chattering neuron, when these neurons are 
connected with excitatory coupling, the networks can exhibit
both tonic firing and network induced bursting with or without noise.   
We will focus on the situation where the networks are bursting
as this is where the mean-field systems can lose accuracy.
The results of simulations using the intrinsically bursting and chattering
neuron parameter values are shown in Figure \ref{fig1}.   
In the bursting region, the frequency error present in the mean-field system is dramatically reduced in the moment-closure reduced PDE, as shown in Figure \ref{fig1}.  Similar results were found for the CA1 and rapidly spiking parameter
values (not shown). This demonstrates that the bulk of the frequency error in the mean-field system is actually due to the separation of time scales approximation.  Thus, the PDE system is superior to the mean-field system in predicting the steady state and dynamics for the actual network.  

To quantify the amount of synchrony in the network, one can use an order parameter defined by : 
\begin{equation}
r(t) = \frac{1}{N}\sum_{i=1}^N \exp\left(2\pi i \left[\frac{v-v_{reset}}{v-v_{peak}}\right]\right) 
\label{orderparam}
\end{equation}
which has been done for example in \cite{AbbottandVreeswijk}.  If $|r(t)| =1$, then the neurons are perfectly synchronized across the network, while if $|r(t)|=0$, they are asynchronous, with the $z_i$ uniformly distributed around the unit cycle.  
As shown in Figures \ref{fig1}(b) and \ref{fig1}(d), the first order moment closure equation provides a great deal more information about synchrony than the mean-field system.  

In addition to the plain Izhikevich model derived from topological normal form theory, various modifications have been suggested to make model better fit the  spiking dynamics and spike profiles for different neurons.  For example, the model can be fit to a fast spiking inhibitory interneuron via the following (see page 299 of \cite{IzBook})
\begin{equation}
\dot{w} = \begin{cases} a((v-v_{b})^3 - w) & \text{if} \quad v\geq v_{b} \\ -aw& \text{if}\quad v<v_{b}\end{cases} 
\label{nlweq}
\end{equation}
Additionally, it is possible to fit sharper spike upstrokes present in actual neurons via the following adjustment: 
 \[
\dot{v} = k(v)v(v-\alpha) - w + gs(e_r-v) + I \\
\]
where
\begin{equation}
k(v) = \begin{cases} k_{min} & \text{if} \quad v\leq \alpha \\ 1 & \text{if} \quad v>\alpha \end{cases}
\label{kswitching}
\end{equation}
This has been done for a hippocampal CA3 pyramidal neuron in \cite{us} in addition to other examples in \cite{IzBook}. The parameter values for these models are given in Table~\ref{table_param}.

For both of these modified Izhikevich models, one can derive the 
corresponding moment-closure reduced PDE and mean-field system.
Comparisons of simulations of these systems with those of the full network are shown in Figure \ref{fig2}.  It is clear that in both cases, the PDE substantially outperforms the mean-field system, both in reproducing network behaviour and
capturing synchrony levels. 

\section{Stability Analysis and Transition to Bursting}\label{stability}

As discussed above and studied in several papers \cite{us2,us3,nesse08}
an important phenomenon of the network behaviour is the transition from 
tonic firing to bursting. In particular, we may wish to characterize how 
this transition depends on various parameters in the model. In principle 
this can be done by running many simulations of the model, but this can 
be time consuming, thus in this section we will explore how we may use the 
reduced models derived in the previous section to do this characterization.

To begin, we generated some benchmark examples using simulations of the 
full model \eqref{mod1}-\eqref{mod3}.  We simulated the network over 
a mesh of values of the parameters $g$ and $I$ for several values of $\sigma$.  
This is shown, by the magenta curves, for networks of chattering neurons in 
Figure \ref{biffig}(a) and intrinsically bursting neurons in Figure \ref{biffig}(b).  
For both these parameter sets (and others not shown), the general bifurcation 
diagram is as follows: without noise, above rheobase ($I_{rh}$) there is an enclosed bursting 
region surrounded by a tonic firing region, while below rheobase there is quiescence.  
Once noise is added, both the bursting region and the tonic firing 
region extend below rheobase, dramatically altering the dynamics of the network.  
Thus, these simulations suggest the network can exhibit noise induced bursting.

Since the mean-field model is a system of ODEs, it can be studied using
numerical bifcuration analysis. In fact, 
numerical two-parameter bifurcation analysis of the mean-field system for
the noiseless network was done in \cite{us2} to study the emergence of 
bursting in the network. It was shown that the bifurcation to bursting for 
$I>I_{rh}$ is via a non-smooth saddle-node of limit cycles closely 
associated with a smooth sub-critical Hopf bifurcation of the tonic firing
equilibrium point. Since the two 
bifurcations occurred closely together, it was found that, for $I>I_{rh}$, 
the Hopf bifurcation 
curves for the mean-field system were a good predictor of the boundary of the 
bursting region for the full network.  The Hopf bifurcation curves are easier
to obtain numerically since they can be found using standard numerical
continuation packages such as MATCONT \cite{matcont}.
Motivated by this work we used MATCONT to find the Hopf bifurcation curves 
for the mean-field system of the network with noise, eqs. 
\eqref{mf1}-\eqref{mf2} (see Appendix B for numerical details).
These curves are shown in Figure \ref{biffig} (see figure caption for details) 
The agreement with the bursting regions for the full network is good. In 
particular, the mean field model for the networks reproduces the fact that, 
in the presence of noise, the Hopf bifurcation curves self-intersect and form regions which extend below $I=I_{rh}$. 

However, there are discrepencies between the mean-field results and the full 
network simulations. Thus, motivated by the results of the previous section, 
we will attempt to use the information in the moment-closure PDE system to 
improve these results.  Our approach will be to study the stability of the 
asynchronous tonic firing solution, since its loss of stability is closely 
associated with the transition to bursing.


The class of models considered present challenges when attempting to analyze the full probability density function eqs. (\ref{pde1})--(\ref{syn_eqn}).  The stability of the 
steady-state solution with $\sigma>0$, or asynchronous state, is commonly analyzed by linearizing the nonlinear partial differential equation around this solution 
\cite{strogatz1991sip,AbbottandVreeswijk,brunelhakim}.  The original system has a two dimensional PDE: while two dimensional PDEs are often tractable, the equations, and in 
particular the boundary conditions (\ref{bc1})--(\ref{bc2}) for these networks, present numerical and analytical difficulties that are not easily resolved with standard methods.  Although the entire 
spectrum of eigenvalues of the linearized PDE contains abundant information about the (infinite-dimensional) system, we choose to consider a lower dimensional subset of variables that is still 
insightful.  Since the population firing rate $\nu(s,\langle w \rangle;\sigma)$ feeds into {\it both} the synapse variable $s(t)$ and the (mean) adaptation variable $\langle w(t)\rangle$, we will analyze the stability of 
the steady-state values of these two variables \cite{nesse08}.

Omitting some details of the PDE, the first order moment closure approximation to the system can be rewritten as:
\begin{eqnarray} 
	\dot{\langle w \rangle} &=& \frac{\langle W_\infty( v )\rangle - \langle w\rangle}{\langle\tau_W( v)\rangle} + w_{jump}\nu(s,\langle w \rangle) =  W(s,\langle w \rangle) \label{w_dyn}\\
	\dot{s} &=& -\frac{s}{\tau_s} + s_{jump}\nu(s,\langle w \rangle)=S(s,\langle w \rangle)\label{s_dyn}  \\
\rho(v,\langle w \rangle,s) &=& \nu(s,\langle w \rangle)\int_v^{v_{peak}}\exp\left(-\frac{2}{\sigma^2}(M(v',\langle w \rangle,s)-M(v,\langle w \rangle,s\right)\,dv' 
\end{eqnarray}
where 
\begin{equation} 
\nu(s,\langle w \rangle) = \left(\frac{2}{\sigma^2}\int_{v_{reset}}^{v_{peak}} \int_{v'}^{v_{peak}}\exp\left(-\frac{2}{\sigma^2}(M(v'',\langle w \rangle,s)-M(v',\langle w \rangle ,s))\right)dv''dv'\right)^{-1} =: F(s,\langle w \rangle)\label{condr}
\end{equation}
Note that $\langle W _\infty(v)\rangle$, and $\langle \tau_w (v)\rangle $ also depend on $s $ and $\langle w \rangle$ through $\rho(v,\langle w \rangle,s)$.  

We denote the steady-state values of this system by $(\bar{w},\bar{s},\bar{\nu},\bar{v})$; we emphasize that $\bar{w}$ and $\bar{v}$ are the steady-state mean values.  The steady-state solution satisfies:
\begin{eqnarray} 
	\bar{w} &=& \overline{\langle W_\infty({v})\rangle} +\overline{\langle \tau_W(v)\rangle}w_{jump}\bar{\nu} \\
	\bar{s} &=& \tau_s s_{jump} \bar{\nu} \label{sbar_eqn} \\
	\bar{\nu} &=& F(\bar{s},\bar{w}) \\
	\bar{v} &=& \int v \rho_V(v;\bar{w},\bar{s}) \,dv
\end{eqnarray}
For numerical simplicity, we focus only on the first two variables: $(\bar{w},\bar{s})$.  Additionally we restrict the dynamics of $\langle w \rangle$ to the case where $\tau_W(v) = \tau_W$ and $W_\infty(v) = bv$.  Linearizing around the steady state via substituting $(\bar{w},\bar{s})^T + \varepsilon \vec{x}e^{\lambda t}$ yields:
\begin{equation}\label{linstab_syst}
	\frac{d \vec{x}}{dt} = M \vert_{(\bar{w},\bar{s},\bar{\nu},\bar{v})} \vec{x}
\end{equation}
where
\begin{equation}\label{stab_matrix}
	M = \begin{pmatrix} -\frac{1}{\tau_w}+\frac{b}{\tau_w}\frac{\partial \langle v \rangle}{\partial \langle w \rangle} +w_{jump}\frac{\partial \nu}{\partial \langle w\rangle} & \frac{b}{\tau_w}\frac{\partial \langle v \rangle}{\partial s} +w_{jump}\frac{\partial \nu}{\partial {s}} \\s_{jump}\frac{\partial \nu}{\partial s}  & -\frac{1}{\tau_s}+s_{jump}\frac{\partial \nu}{\partial s} \end{pmatrix}.
\end{equation}
The eigenvalues of $M$ indicate the stability of the asynchronous state. 

The stability analysis described thus far is fairly standard.  However, the rest of the calculations described below are different and novel to the best of our knowledge.  

\subsection{An accurate approximation to the steady-state firing rate}\label{section_augmentRate}

We first describe how to calculate the steady-state firing rate $\bar{\nu}$.  Normally, one would use eq. (\ref{nu1}) to calculate $\nu(\langle w \rangle,s)$.  However, in the first 
order moment closure system, much of the information about the density in $w$ is lost in the approximation process, which contributes to the error in the mean-field approach.  Our approach to rectify 
this involves calculating the firing rate in an alternative way to take into account the information about the marginal density in $\rho_w(w)$.  

The rate is calculated via a dimension reduction method based on \cite{Ly_PrincDimRed_13} (also see \cite{moreno1,moreno08,nesse08} for similar approaches) where only 1 dimensional PDEs need to be numerically solved.  
A standard application of the analogous dimension reduction method assumes $w$ is a parameter rather than a random variable and $\bar{s}$ is given (eq. (\ref{sbar_eqn})), 
thus resulting in a 1 dimensional PDE for the steady-state marginal voltage density $\rho_V(v;w,\bar{s})$ 
(eq. (\ref{rho1}) but with $(w,s)$ as parameters).  So we have 
a family of $\rho_V(v;w,\bar{s})$ that depends on $(w,\bar{s})$, which also has a corresponding family of steady-state firing rates that depend on $(w,\bar{s})$: 
$${\tilde \nu}(w,\bar{s}) = F(w,\bar{s})$$ 
where $F(w,s)$ is given by eq. (\ref{condr}).  Again, we interpret the firing rate $\nu(w,\bar{s})$ as a conditional firing rate, conditioned on the variables $(w,\bar{s})$.  
There is a $w$ variable for each individual neuron in the population, with a marginal $w$ density (recall eq. (\ref{pde1})): 
$$\rho_W(w,t):=\int_{v_{reset}}^{v_{peak}}\rho(v,w,t)\,dv.$$  
Solving for the actual $\rho_W(w,t)$ function is difficult numerically and high dimensional, so we make the following approximation for the steady-state $\rho_W(w)$ equation:
\begin{equation}\label{marg_w_approx}
	0 = -\frac{\partial}{\partial w} \left( \frac{\overline{\langle W_\infty(v)\rangle}-w}{\overline{\langle \tau_W(v) \rangle}}\rho_W(w)+\bar{\nu}\int_{w-w_{jump}}^w\rho_W(w')\,dw' \right)
\end{equation}
where the angular brackets in $\langle W_\infty(\bar{v})\rangle$ and $\langle \tau_W(\bar{v}) \rangle$ represent integrating over $\rho_V(v;\bar{w},\bar{s})$.  
This essentially assumes that the firing rate of the population is a Poisson process and the jumps in $w$ are independent of $w$ \cite{ntfast}.  
We can use this approximation for the marginal $w$ density to calculate the population firing rate:
$$\bar{\nu} = \int_0^\infty J(v_{peak};w',\bar{s}) \rho_W(w')\,dw' = \int_0^\infty {\tilde \nu}(w',\bar{s}) \rho_W(w')\,dw'$$
Note that the (average) synapse variable $\bar{s}$ is exactly the same for all neurons, and its steady-state value will be determined by $\bar{\nu}$ (eq. (\ref{sbar_eqn})).  Since this is a nonlinear 
system, the steady-state solution using this reduction method should satisfy the following system of equations:
\begin{eqnarray}
	\bar{s} &=& \tau_s s_{jump} \bar{\nu} \label{selfcon1} \\
	\bar{\nu} &=& \int_0^\infty {\tilde \nu}(w',\bar{s}) \rho_W(w')\,dw' \label{selfcon2}
\end{eqnarray}
Unfortunately, requiring this system to be solved self-consistently predominately results in an unstable system with iteration methods, even when Monte 
Carlo simulations of the true system have very stable asynchronous states and even when $\bar{\nu}$ is set to be the 'correct' value.  Hence, it would seem that applying this approach results 
in instabilities in the numerical solutions.

To rectify this issue, we relaxed the self-consistency condition and consider eqs. (\ref{selfcon1})--(\ref{selfcon2}) as a linear input ($\nu_{in}$) / output ($\nu_{out}$) system or mapping.  Specifically, 
$\nu_{in}$ is used in eq. (\ref{marg_w_approx}) in place of $\bar{\nu}$ to solve for $\rho_W(w;\nu_{in})$; $\nu_{in}$ determines $\bar{s}$ in eq. (\ref{selfcon1}), and that $\bar{s}$ is used in the 
equation for the family of $\rho_V(v;w)$ and thus ${\tilde \nu}(w,\bar{s})$.  Finally, $\nu_{out}=\int_0^\infty {\tilde \nu}(w,\bar{s}) \rho_W(w'; \nu_{in})\,dw'$.  We calculate $(\nu_{in},\nu_{out})$ 
on a fine grid of reasonable $\nu_{in}$ values and select the one with the smallest difference $|\nu_{out}-\nu_{in}|$ as $\bar{\nu}=\nu_{out}$.  This approach is numerically the closest 
approximation to the self-consistent solution for $\nu_{in}=\nu_{out}$.  It turns out this system always has a unique minimum 
$|\nu_{out}-\nu_{in}|$ for the parameters considered, and the approximation to the steady-state firing rate is extremely accurate.

To summarize, we view the nonlinear system as a mapping of $\nu_{in}$ to $\nu_{out}$ with the following sequential steps:
\begin{eqnarray} 
\bar{s} &=& \tau_s s_{jump} \nu_{in}  \label{sc1}\\
0 &=&  -\frac{\partial}{\partial v}\left((F(v)-w+g\bar{s}(er-v)+I)\rho_V-\frac{\sigma^2}{2}\frac{\partial \rho_V}{\partial v} \right),   \text{  calculate a family of }\rho_V(v;w,\bar{s})\\
\nu(w',\bar{s}) &=& J(v_{peak};w',\bar{s}), \text{      and thus, family of firing rates [$J$ is a linear functional of $\rho_V$]}\\
0 &=& -\frac{\partial}{\partial w} \left( \frac{\overline{\langle W_\infty(v)\rangle}-w}{\overline{\langle \tau_W(v) )\rangle}}\rho_W(w)+\nu_{in}\int_{w-w_{jump}}^w\rho_W(w')\,dw' \right), \text{         calculate $\rho_W$}\\
\text{where } & &\overline{\langle W_\infty(v)\rangle} = \int W_\infty(v) \rho_V(v;w=\tau_Ww_{jump}\nu_{in},\bar{s})\,dv \text{   and similarly for $\overline{\langle \tau_W(v) )\rangle}$} \\
\nu_{out} &=& \int_0^\infty \nu(w',\bar{s}) \rho_W(w')\,dw' \label{scend} 
\end{eqnarray}
Recall that the steady state values corresponds to minimizing $|\nu_{out}-\nu_{in}|$.

We remark that this method is {\bf not} the same firing rate from the first order moment closure equations (\ref{sys1})--(\ref{sys3}) because in those equations only the mean of the $w$ variable is used, 
not its probability distribution.  In particular, in the traditional mean-field approach, one uses $\nu(s,\langle w \rangle)$ and $\rho_V(v,\langle w \rangle,s)$ and interprets these as the network averaged 
firing rate, and the marginal density in $v$ as simple functions of $\langle w\rangle$.  In our approach, we use the same equations but now interpret $\nu(s,w)$ and $\rho_V(v;w,s)$ as 
the conditional quantities (conditioned on $w$), and use a pragmatic approximation to the self-consistency condition.  This allows us to incorporate information about the density in $w$ with an approximation 
for the marginal $w$ density.

\subsection{Linear stability analysis with approximation to steady-state firing rate}

With the approximation method for the steady-state firing rate in section \ref{section_augmentRate} summarized in eqs. (\ref{sc1})-(\ref{scend}), we can numerically perform the stability analysis described in equations 
(\ref{linstab_syst})--(\ref{stab_matrix}) assuming that $w(t)$ represents the population average.  Note that this stability analysis of a two-dimensional nonlinear 
PDE system does not rely on any Monte Carlo simulations, but rather just analyses and 
reductions based on the PDEs.  In the matrix in eq. (\ref{stab_matrix}), the partial derivatives of the steady-state firing rate with respect to $\bar{s}$ and $\bar{w}$ are calculated numerically using a finite 
difference method.

This method is implemented for the Izhikevich all-to-all neural network for two parameter sets: chattering neurons (Figure \ref{CHbif}, black dotted curve) and intrinsically bursting neurons (Figure \ref{IBbif}, black dotted curve).  
Since this method is non-standard, we were not able to leverage MATCONT \cite{matcont} to numerically continue the bifurcation points but rather had to perform the analysis on a fine grid in parameter space.  Over a two 
dimensional region of parameter space where the behavior varies appreciably, an implementation of the method is able to capture the regions where the neural network exhibits oscillations and quiescence 
(black dotted curves in Figure \ref{biffig}).  The feature of noise-induced bursting is also captured with the method, as well as the qualitative shape of the various regions of stability.  
We omit the curve for $\sigma=0$ because a standard discretization of the operators requires 
manual refinement of the various meshes and is quite tedious; note that the black dotted curves are for a fixed discretization using a standard finite difference method.  
Furthermore, despite only focusing on two variables in the system, this approach gives an approximation to the marginal voltage density $\rho_V(v)$ that matches well with the Monte Carlo 
simulations (not shown), and as already mentioned it also provides an approximation to the steady-state firing rate, and marginal $w$ density.  

\section{Discussion}\label{disc}

We considered a population density approach to study the dynamics of large networks of integrate-and-fire
type models with adaptation. We presented a first order moment closure reduction which results in a one 
dimensional partial differential equation for the density of the voltage, $\rho_V(v,t)$, coupled to a
two dimensional system of ODEs for the network mean adaptation and synaptic activity.  We obtained an analytical
solution for the steady-state voltage density and use this to derive a steady-state mean-field system
for the network.  When applied to various recurrently coupled spiking networks, the PDE-ODE system is able 
to successfully capture a large range of transient dynamics of the network.  
In contrast to a steady-state mean-field system, the frequency error in capturing oscillations, i.e., bursts, is reduced if not absent in the coupled PDE-ODE system.  Additionally, one obtains information about synchrony and other rapid temporal correlations with the reduced population density equations, unlike in the mean-field system.  However, one can still use the mean-field approach for a white noise system to ascertain the stability of the steady states and slow oscillations as before.

A novel linear stability analysis method was presented and applied to 
particular instances of these class of neural network models.  The method is also able to predict the bursting region for the network of neurons.  The method has a pragmatic solution for dealing with a dimension reduction method that 
would make a bad problem worse (see \cite{ly_tranchina_07} for similar issues with higher order moment closure methods), and does provide approximations for other entities of interest (marginal densities, firing rate).  
However, the method is impractical in leveraging continuation software \cite{doedel86,matcont} currently and would 
likely require more programming and development to do so.  Even though the dimension reduction method and corresponding linear stability analysis could be applied to other systems, the details of the implementation could present 
specific technical and numerical challenges in itself.  Nevertheless, taken together these results are valuable and will hopefully be insightful for other nonlinear systems with higher dimensional PDEs 
that require dimension reduction.

Brunel \& Latham \cite{BL03} analyzed a two-dimensional quadratic integrate-and-fire with temporally correlated noise, and calculated the population firing rates in 
various regimes.  Their state variables were voltage and the noise (Ornstein-Uhlenbeck) forcing, and did not include an adaptation variable.  Their work resulted in analytic formulas for the 
firing rates in the slow and fast colored noise limits.  They suggest using the mean firing rate for the purposes of a mean-field system, and indeed we do apply their idea here.  However, the networks in \cite{BL03} were uncoupled, and non-adapting quadratic integrate-and-fire neurons.  Thus, the network cannot display bursting as the intrinsic dynamics of the neurons do not support bursting at the individual level, and without coupling the network cannot display emergent bursting at the network level.  Also, it appears that for accurate estimates of the frequency of bursting, a mean-field system is not sufficient and one has to numerically solve at least the marginal voltage density to obtain the correct dynamics. 

Similarly, Richardson \cite{richardson07,richardson08} considered nonlinear integrate-and-fire networks (e.g., exponential and Izhikevich) without adaptation, where 
analytic formulas were provided for various network statistics.  In particular, the firing rate quantities \cite{richardson07} and the spike train spectra and first passage time density \cite{richardson08} were calculated.  Adaptation 
has been considered in the context of noisy nonlinear neural networks for example by \cite{richardson09}.  In that paper, the author considered recurrent noisy nonlinear integrate-and-fire networks with biophysical adaptation currents using a 
similar Fokker-Planck or population density formalism.  Their analyses were based on linear response theory with small amplitude sinusoidal drive and relied on a separation of time scales between the two state variables.  
Thus, our work differs from 
\cite{richardson09} not only in the functional forms of the equations, but also because our stability analyses were different (first order moment closure and the method described in section \ref{section_augmentRate}).  
To the best of our knowledge, they did not consider oscillatory or bursting regimes; oscillatory firing in their work appears to be primarily driven by background sinusoidal inputs.  

In the work of \cite{nesse08}, a network of leaky integrate-and-fire neuron with noise was studied with a mean-field model using bifurcation analysis.  
The mean-field system they derive is very different from the one considered here. 
In particular, the noise in their system is synaptically filtered through a double exponential synapse, 
so the correlation time in the noise is fairly high and can be treated as static heterogeneity.  
Using $I$ and $\sigma$ as the bifurcation parameters, they show that transitions to bursting occur via both 
subcritical and supercritical Hopf bifurcations 
and that co-dimension 2 Bautin points occur at the interface between these two kinds of bifurcations.  Recently, we showed in \cite{us3} that the bifurcation sequence in a static heterogeneous network is 
identical to the one in \cite{nesse08}, aside from the model differences.  We remark that there are complications that arise with the notion of a mean-field system for a heterogeneous network of neurons, as 
there is no unique mean-field system in this case because multiple systems can be derived depending on what assumptions are used \cite{us3}.  
Thus, there is still some insight to be gained by analyzing the network/mean-field system in the true white noise limit, as opposed to the large time 
correlation limit in the correlation function for the noise. 

While the mean-field system we derive does have some error in terms of the dynamics of the network 
level oscillations, it appears to be quite accurate for the steady-state firing rate and its stability.  
In particular, it shows that the region of bursting, which lies completely in the $I>I_{rh}$ part of
parameter space in the noiseless case, extends below $I=I_{rh}$ when noise is present. Further, the
curve of Hopf bifurcations associated with the emergence of bursting becomes a self-intersecting
curve in the presence of noise.  Recent analytical work \cite{us4} in the noiseless case, has shown 
there is a region of coexistence of quiescence and tonic firing when $I<I_{rh}$ 
and that there are several non-smooth co-dimension 2 bifurcations on 
$I=I_{rh}$ which are associated with the loss of bursting for $I<I_{rh}$.  
Preliminary numerical bifurcation analysis of the mean-field equations 
indicates that the non-smooth co-dimension 2 bifurcations found in the
noiseless case become regularized as smooth co-dimension 2 bifurcations
when noise is added.  
The emergence of bursting for $I<I_{rh}$ and the associated change in the 
Hopf bifurcation curves is worthy 
of further analysis, but beyond the scope of this paper.  
We leave futher investigation of these bifurcations
for future work.



For the full population density equations, a future direction may be to numerically solve the two-dimensional partial differential equation(s) coupled with the ordinary differential equations.  Standard finite 
difference methods, even with higher orders of accuracy, proved to be unstable and not accurate compared to Monte Carlo simulations.  Therefore, developing the numerical 
solutions to these equations is nontrivial and beyond the scope of this paper, but has the potential benefit of capturing the full statistical quantities of the network and may provide further analytical 
insights.  To our knowledge, only a rudimentary first order method appears in the literature for this system.

\begin{table}
\centering
\begin{tabular}{|c|c|c|c|c|c|c|}
\hline
Parameter Set & CA1 & CH & IB & RS & FS & KS \\ 
\hline \hline
$\alpha$ &  0.25& 0.33 & 0.4 & 0.33 & 0.18 &0.72 \\
\hline
$v_{reset}$ & 0.25 & 0.33 & 0.25 & 0.17 &0.18&0.154\\
\hline
$v_{peak}$ & 1.67 & 1.42 & 1.67 & 1.58& 1.45&1.462\\
\hline
$w_{jump}$ & 0.028 & 0.028 & 0.019 & 0.04 &0 &0.012\\
\hline
$1/\tau_w $ & 0.033 & 0.017 & 0.017 & 0.07& 0.2&0.005 \\
\hline
$b$ & 0.017 & 0.011 & 0.056 & -0.048 &1.38 &-0.003\\
\hline
\end{tabular}
\caption{The dimensionless parameters for the fitted Izhikevich models used in network, mean-field, and population density simulations throughout the text.  
The models are CA1 pyramidal cell (CA1), chattering neuron (CH), intrinsically
bursting neuron (IB), rapid spiking neuron (RS), fast spiking (FS) and
k-switching (KS).  The corresponding dimensional parameters can be found 
in \cite{IzBook}.  The values of the following
parameters were the same for all simulations:  
$\tau_s = 1.5, e_r = 1$, and $s_{jump} = 1$.  The parameters $g$ and $I$ 
and $\sigma$ vary, and are treated as bifurcation parameters. }\label{table_param}
\end{table}

\renewcommand{\abstractname}{Acknowledgements}
\begin{abstract}
This work benefitted from the support of the Natural Sciences and Engineering Research Council of Canada and the Ontario Graduate Scholarship program
\end{abstract}

\section*{Appendix A:  The Extended Mean-Field System} 
One can also apply the methods in developed this paper on the extended interval, $(-\infty,v_{peak}]$.  In particular, the one-dimensional moment closure PDE has to be solved on a larger interval with the boundary condition (\ref{bc1}) which is easily discretized in space.   For the mean-field system, one has to solve for the steady state $\rho(v)$ and $\nu$ on the interval $(-\infty,v_{peak}]$.   Note that due to the boundary condition, the density function $\rho(v)$ will be continuous at steady state, but not differentiable as the flux is piecewise constant at steady state, and given by: 
\begin{eqnarray}
J(v,s,\langle w \rangle) = \begin{cases} \nu(s,\langle w \rangle)  & v_{reset} \leq v\leq v_{peak}  \\ 0 & v<v_{reset}\end{cases} 
\end{eqnarray} 
Solving for the density function on these two intervals, and forcing continuity of the density function at $\rho(v_{reset})$ yields the following: 
\begin{eqnarray}
\rho(v,\sigma) = \begin{cases}\nu\frac{2}{\sigma^2} \int_{v}^{v_{peak}}\exp\left(-\frac{2}{\sigma^2}\left[M(v',\langle w \rangle,s)-M(v,\langle w \rangle,s) \right]\right)\,dv'  & v_{reset}\leq v \leq v_{peak} \\ \nu\frac{2}{\sigma^2} \int_{v_{reset}}^{v_{peak}}\exp\left(-\frac{2}{\sigma^2}\left[M(v',\langle w \rangle,s)-M(v,\langle w \rangle,s) \right]\right)\,dv' & v<v_{reset}  \end{cases} 
\end{eqnarray} 
In order to determine $\nu$, one has to us the normalization condition on $\rho(v,\sigma)$ to yield: 
\begin{eqnarray}
\nu^{-1} &=&\frac{2}{\sigma^2}\left[ \int_{v_{reset}}^{v_{peak}} \int_{v}^{v_{peak}}\exp\left(-\frac{2}{\sigma^2}\left[M(v',\langle w \rangle,s)-M(v,\langle w \rangle,s) \right]\right)\,dv'dv \nonumber \right.\\
&+& \left.\int_{-\infty}^{v_{reset}} \int_{v_{reset}}^{v_{peak}}\exp\left(-\frac{2}{\sigma^2}\left[M(v',\langle w \rangle,s)-M(v,\langle w \rangle,s) \right]\right)\,dv'dv \right]
\end{eqnarray} 
As in section~\ref{mfsec}, one can use Laplace's method to prove convergence of the mean-field system with noise to the noiseless mean-field system as $\sigma \rightarrow 0$.  

\section*{Appendix B: Implementing the Mean-Field System}  
In order to numerically simulate the mean-field system derived in 
section~\ref{mfsec}, one has to compute the integral: 
\begin{equation}
\nu(\sigma,s,\langle w \rangle)^{-1} = \frac{2}{\sigma^2}\int_{v_{reset}}^{v_{peak}} \int_{v}^{v_{peak}}\exp\left(-\frac{2}{\sigma^2}(M(v',\langle w \rangle,s)-M(v,\langle w \rangle ,s))\right)dv'dv
\end{equation}
as a function of $s$ and $\langle w \rangle$ at each time step.  This requires numerically computing a double integral over a triangular region in the $v$ plane.  As $\sigma \rightarrow 0$, the exponential term inside the integral often becomes difficult to work with due to the $\frac{1}{\sigma^2}$.  However, by using the substituion  $v' = v + \frac{\sigma^2}{2} z$, one arrives at the integral: 
\begin{eqnarray*}
\nu(\sigma,s,\langle w \rangle)^{-1} &=& \int_{v_{reset}}^{v_{peak}} \int_{0}^{\frac{2}{\sigma^2}(v_{peak}-v)  }\exp\left(-\frac{2}{\sigma^2}\left[M\left(v+\frac{\sigma^2 z}{2},\langle w \rangle,s\right)-M(v,\langle w \rangle ,s)\right]\right)dz dv\\
&=& \int_{v_{reset}}^{v_{peak}} \int_{0}^{\frac{2}{\sigma^2}(v_{peak}-v)  }\exp\left(-\left[\sum_{i=1}^\infty \frac{\partial^i M(v,\langle w \rangle,s)}{\partial v^i}z^{i}\left(\frac{\sigma^2}{2}\right)^{i-1}\right]\right)dz dv \\
&=&\int_{v_{reset}}^{v_{peak}} \int_{0}^{\frac{2}{\sigma^2}(v_{peak}-v_{reset})  }\exp\left(-\left[\sum_{i=1}^\infty \frac{\partial^i M(v,\langle w \rangle,s)}{\partial v^i}z^{i}\left(\frac{\sigma^2}{2}\right)^{i-1}\right]\right) H\left(\frac{2}{\sigma^2}(v_{peak}-v) - z \right)dz dv \\
\end{eqnarray*}
Note that the term inside the exponential no longer has a $\frac{2}{\sigma^2}$ term which yields numerical difficulties in the $\sigma \rightarrow 0$ limit.  While the bounds of the integral now diverge as the upper bound now has a $\frac{\sigma^2}{2}$, the integrand converges to zero for large $z$ exponentially fast.  For the Izhikevich and quartic integrate and fire models, there is only a finite number of terms in the sum, as $F(v)$ and thus $M(v)$ is a polynomial in $v$.  For other models, one can take a finite number of terms to approximate the firing rate.  The Heaviside function $H(x)$ converts the triangular integration region into a rectangular one.  The remaining integral can be simply computed with the two-dimensional trapezoidal method over a rectangular region.  The Matlab function trapz is used to compute the integral at each time step over a two-dimensional finite mesh in the $v'$ and $z$ variables.  This is used for both direct simulation of the mean-field system and numerical bifurcation analysis of the system in MATCONT.   Note that this implementation is similar to the one suggested in \cite{BL03} only we compute the firing rate at each time step, as there does not appear to be much computational overhead in this approach versus using a function table, as first suggested in \cite{BL03}.

\section*{Appendix C:  Validity of the First Order Moment Closure Assumption}\label{err}
In the derivation of section~\ref{momclos}, we had to assume independence of specific moments, a series expansion in $w_{jump}$, and the first order moment closure approximation $\langle w | v \rangle  = \langle w \rangle$.  While an error bound on the first assumption is difficult to arrive at, we can show that under certain assumptions, the first order moment closure does not contribute much to the error if $w_{jump}$ is small.  In particular, one can show that the conditional moment $\langle w | v \rangle$ has to satisfy the condition 
\begin{equation}
\langle w | v_{reset}\rangle = \langle w | v_{peak}\rangle + w_{jump}\label{bcw}
\end{equation}
when the network is undergoing firing (for $I>I^*(s,\langle w \rangle)$) if the following conditions hold:
\begin{eqnarray}
\sigma^2_{w|v}\rho_V(v,t)\bigg|_{\partial V} + \frac{\sigma^2}{2}\frac{\partial \langle w|v \rangle}{\partial v}\rho_V(v,t)\bigg|_{\partial V} = 0
\end{eqnarray}
which holds with the higher order moment closure assumption in the $\sigma = 0$ case discussed in section \ref{momclos}. The condition \ref{bcw} can be derived by looking at the differential equation for $\langle w \rangle$ by changing the order of integration ($w$ first then $v$ as opposed to $v$ first then $w$) and equating the two resulting expressions for $\langle w \rangle'$.

\begin{figure}
        \centering
        \begin{subfigure}{0.45\textwidth}
                \centering
                \includegraphics[width=\textwidth]{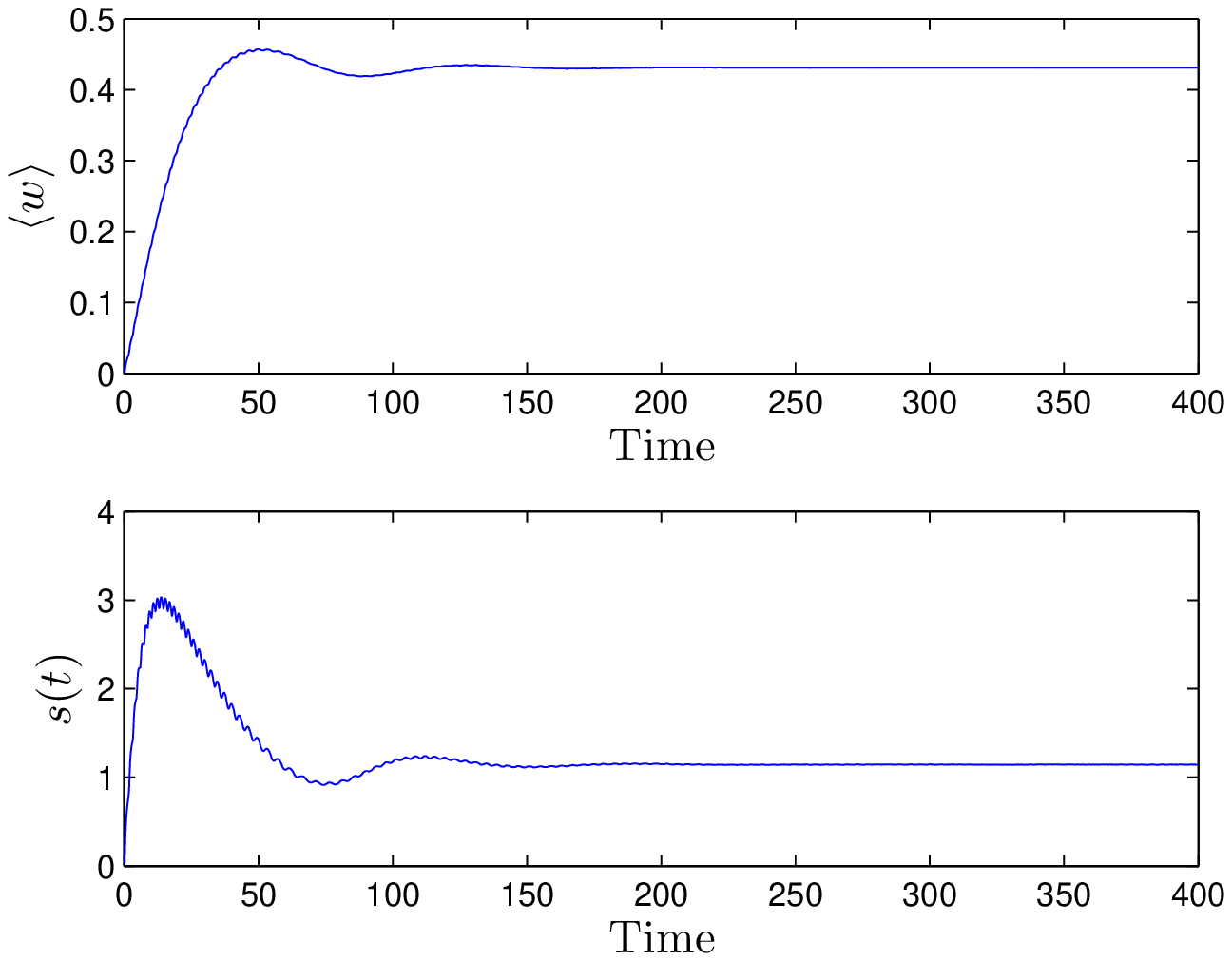}
                \caption{Tonic Firing $s(t)$ and $\langle w \rangle$}   
                \label{figwv1}
        \end{subfigure}
\quad 
        \begin{subfigure}{0.45\textwidth}
                \centering
                \includegraphics[width=\textwidth]{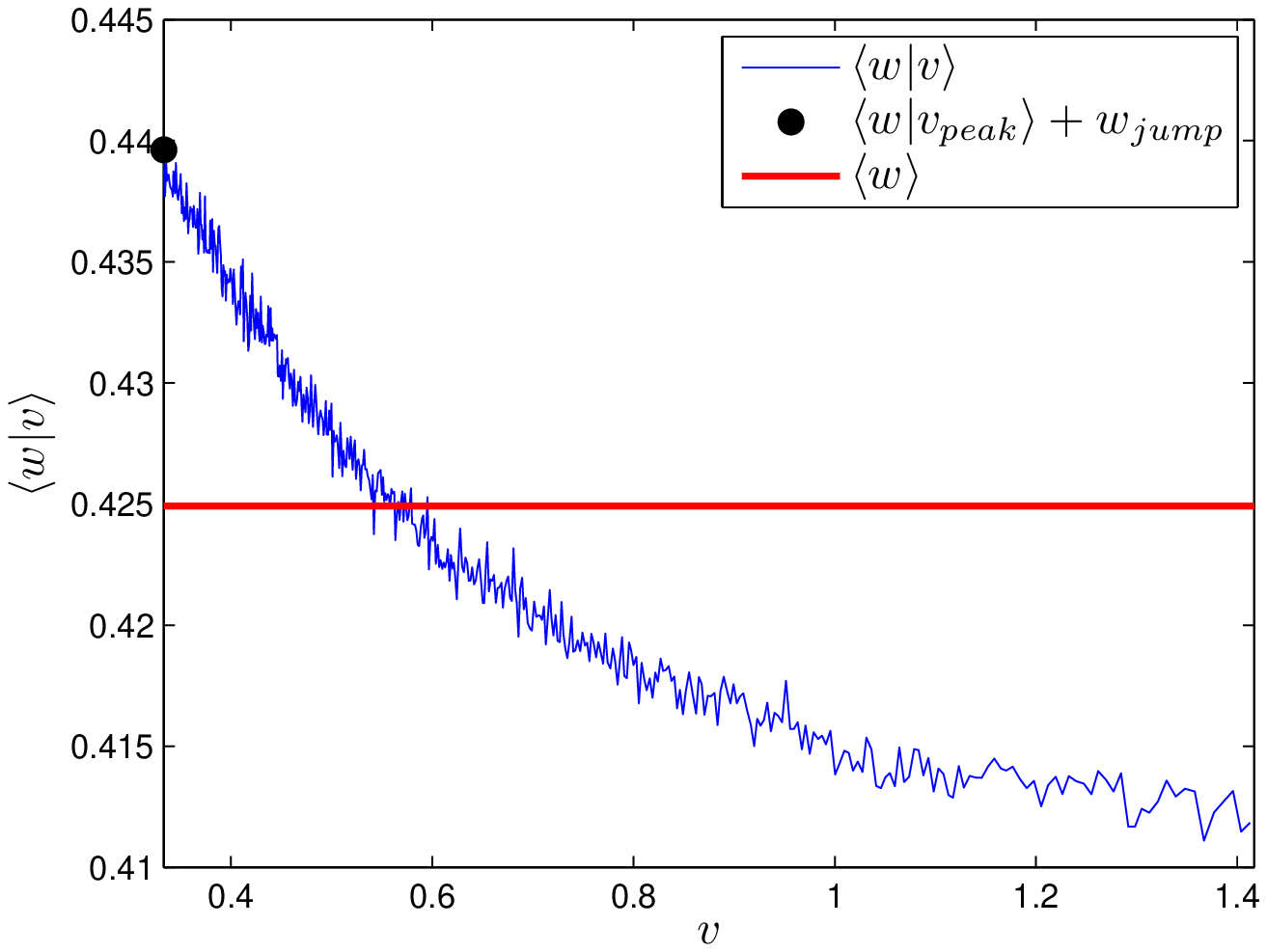}
                \caption{$\langle w | v\rangle$ at $t = 200$}
                \label{figwv2}
        \end{subfigure}\\

        \begin{subfigure}{0.45\textwidth}
                \centering
                \includegraphics[width=\textwidth]{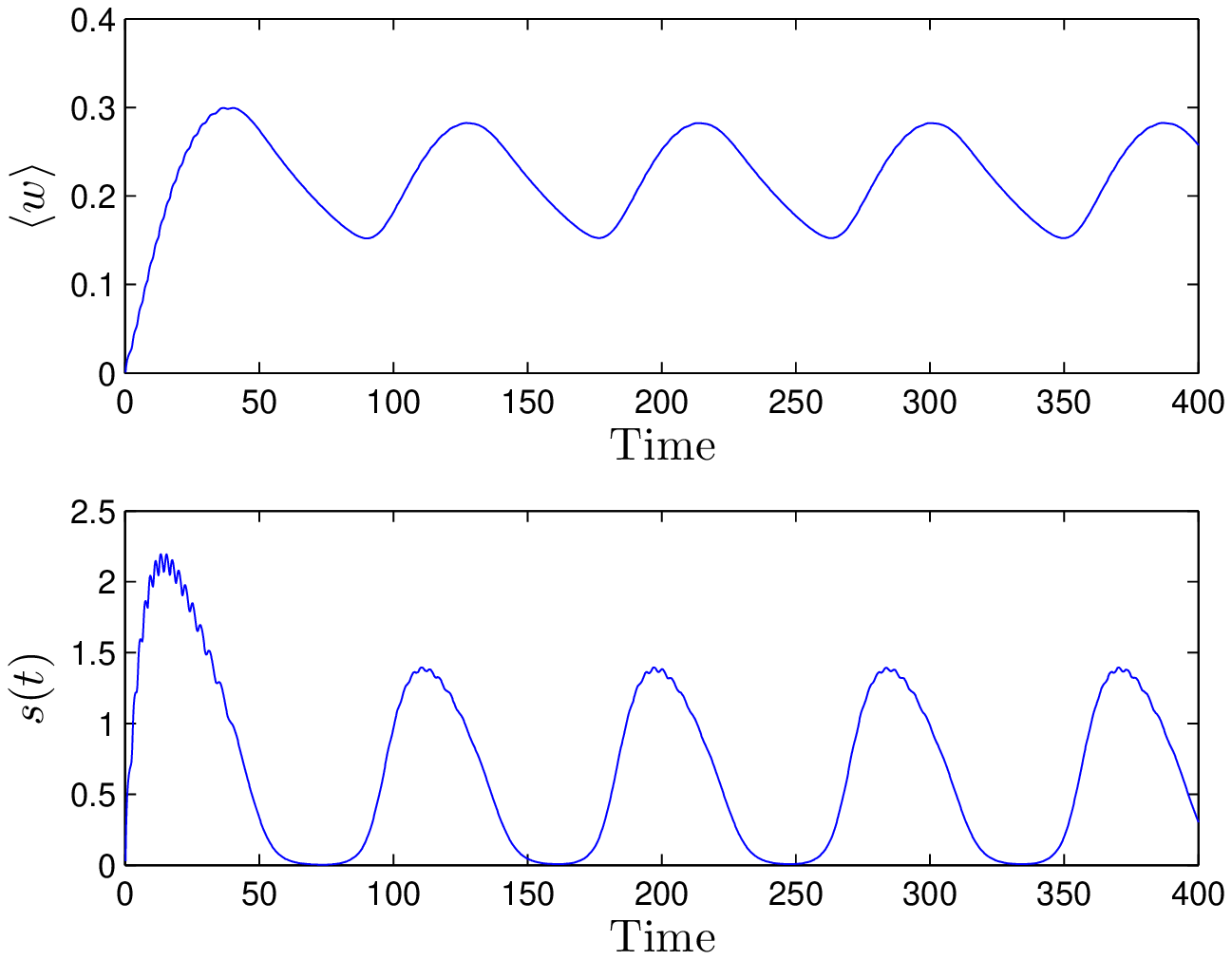}
                \caption{Bursting $s(t)$ and $\langle w \rangle$}
                \label{figwv3}
        \end{subfigure}
        \begin{subfigure}[(b)]{0.45\textwidth}
                \centering
                \includegraphics[width=\textwidth]{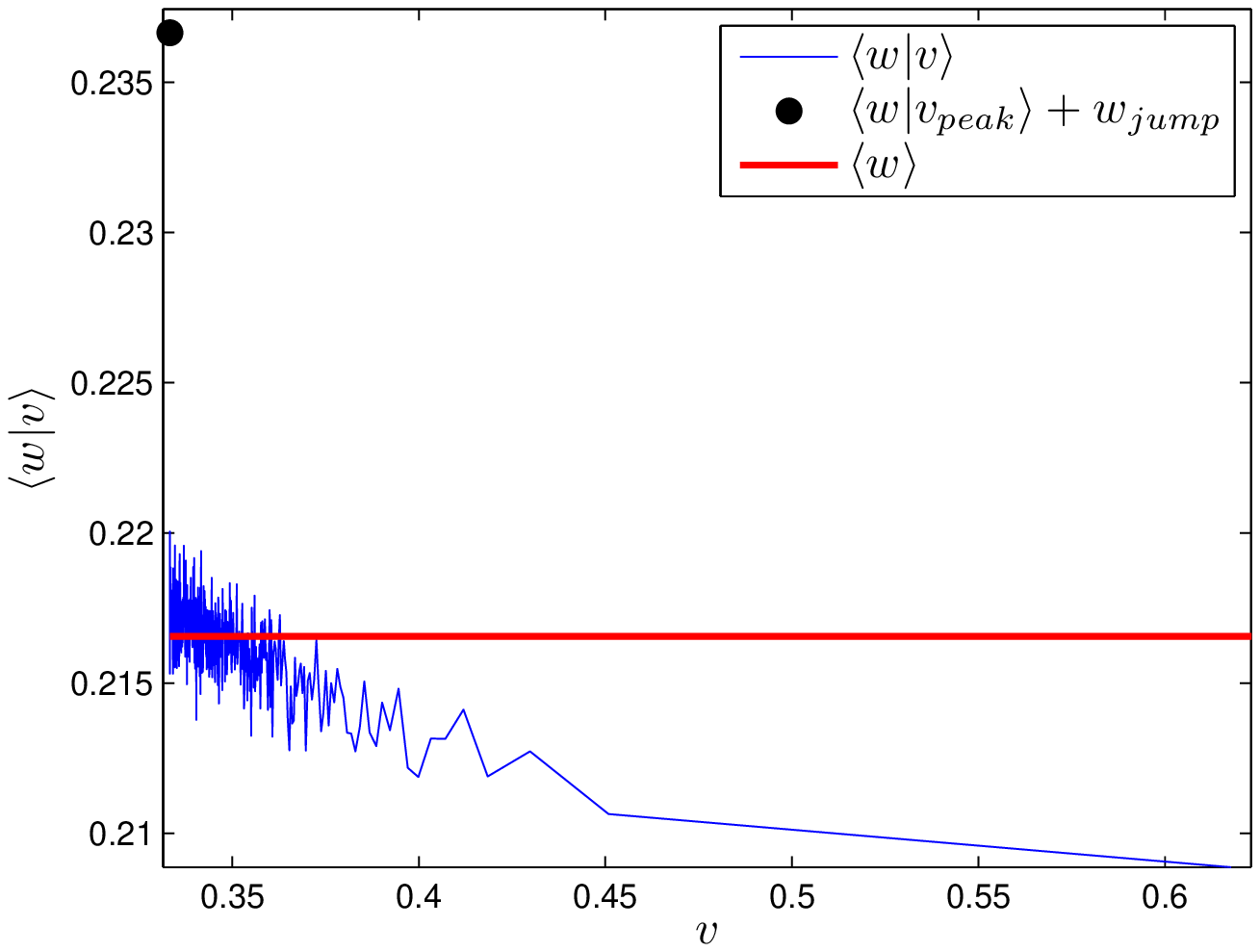}
                \caption{$\langle w | v\rangle$ at $t = 150$.}  
                \label{figwv4}
        \end{subfigure}
\caption{The first conditional moment $\langle w | v\rangle$ is computed by sorting the $w_i$ as a increasing function of $v_i$ and then averaging locally the $v_i$ and $w_i$.  A network of 50,000 neurons was simulated using the chattering neuron (CH) parameter sets in Table \ref{table_param} in either the tonic firing (a)-(b) ($g=0.33$, $I=0.29$, $\sigma = 0.05$) or the bursting regions (c)-(d)  ($g=0.33$, $I=0.11$, $\sigma = 0.05$).  Note that $\langle w | v_{peak}\rangle + w_{jump}$ is plotted at $v=v_{reset}$ (black dot in (b),(d)) to demonstrate the validity of the boundary condition in the tonic firing region.  The red line is $\langle w \rangle$. In both the tonic firing and bursting regions, $\langle w | v\rangle$ is a monotonically decreasing function of $v$ with a narrow range.  When the network is bursting, $\langle w|v_{reset}\rangle = \langle w |v_{peak}\rangle + w_{jump}$ during the active portion of the bursts, and $\langle w | v_{reset}\rangle < \langle w | v_{peak}\rangle + w_{jump}$ during the quiescent periods.    } \label{FIGWV}

\end{figure}

\begin{figure}
        \centering
        \begin{subfigure}[b]{0.47\textwidth}
                \centering
                \includegraphics[width=\textwidth]{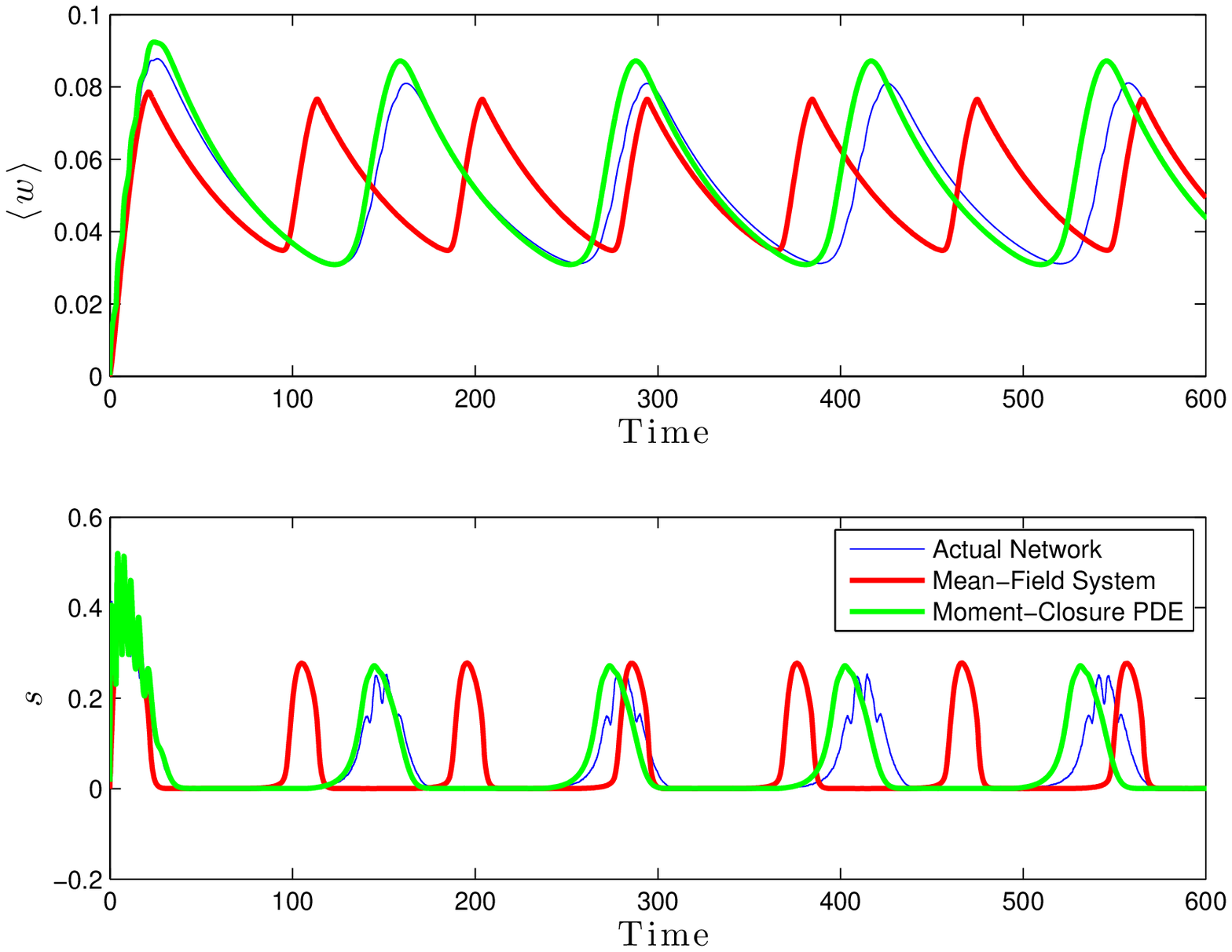}
                \caption{Izhikevich Model, IB Parameters}
                \label{IB1}
        \end{subfigure}
\quad 
        \begin{subfigure}[b]{0.47\textwidth}
                \centering
                \includegraphics[width=\textwidth]{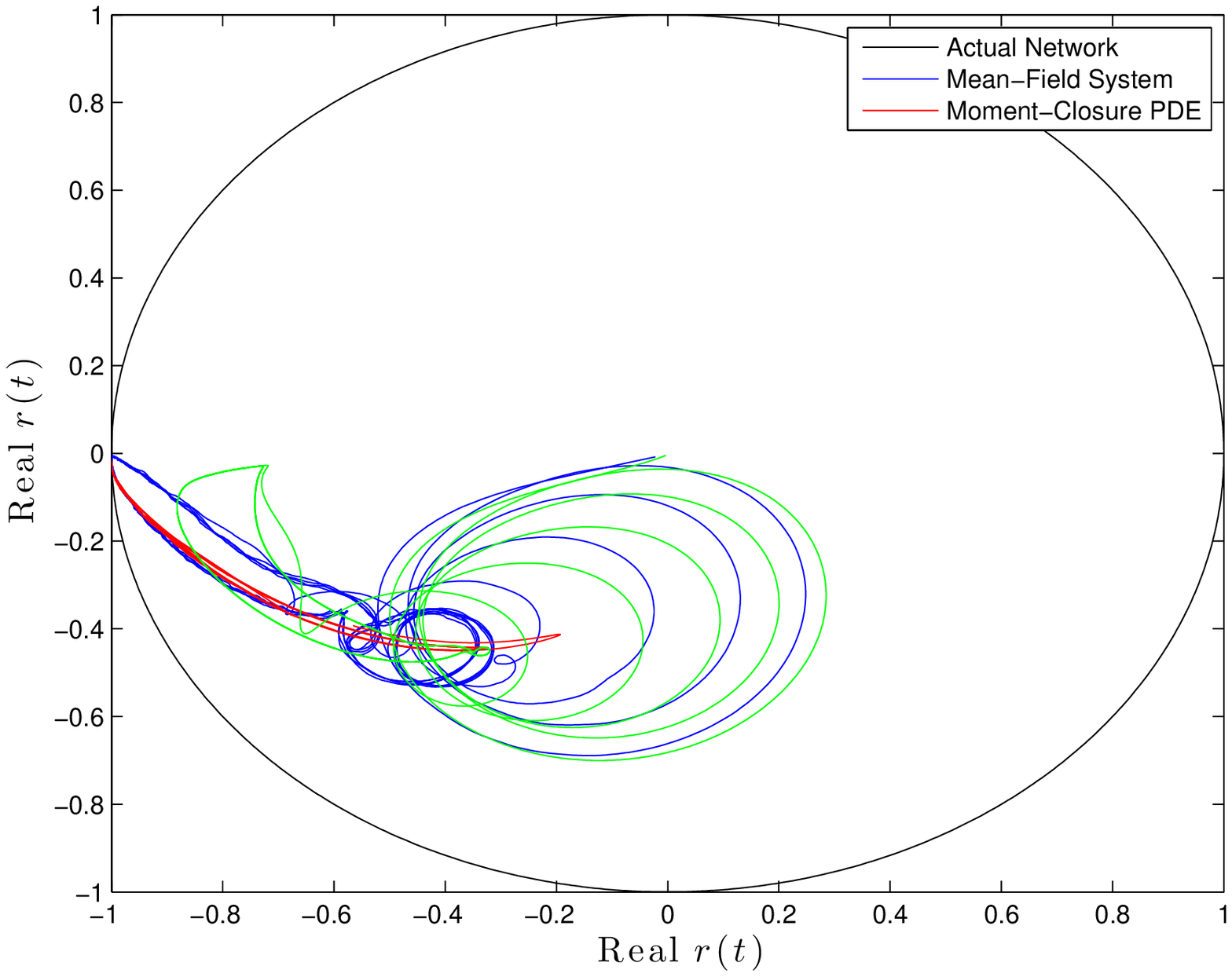}
                \caption{Order Parameter, IB}
                \label{IB2}
        \end{subfigure}
\\
        \begin{subfigure}[b]{0.47\textwidth}
                \centering
                \includegraphics[width=\textwidth]{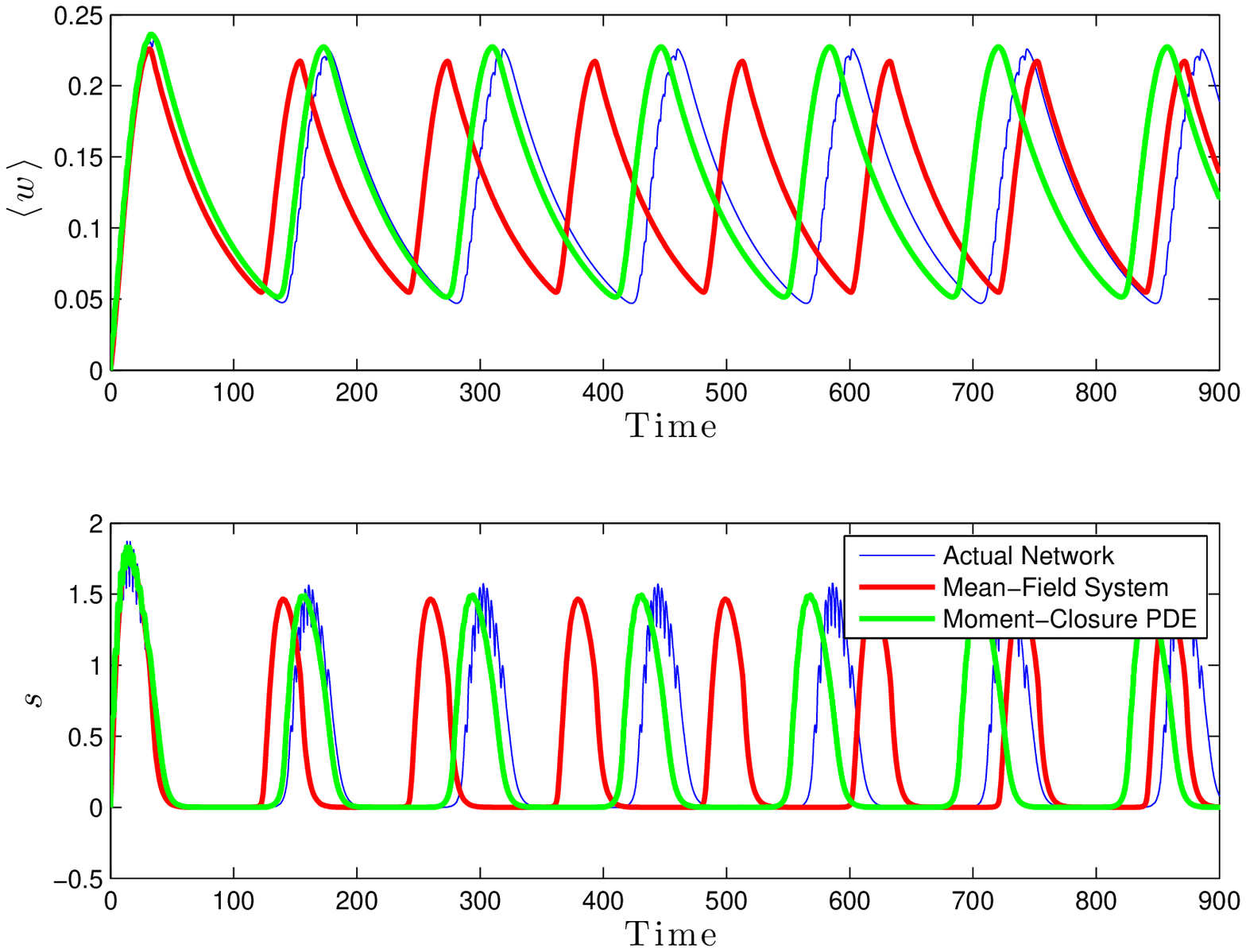}
                \caption{Izhikevich Model, CH Parameters}
                \label{CH1}
        \end{subfigure}
\quad 
        \begin{subfigure}[b]{0.47\textwidth}
                \centering
                \includegraphics[width=\textwidth]{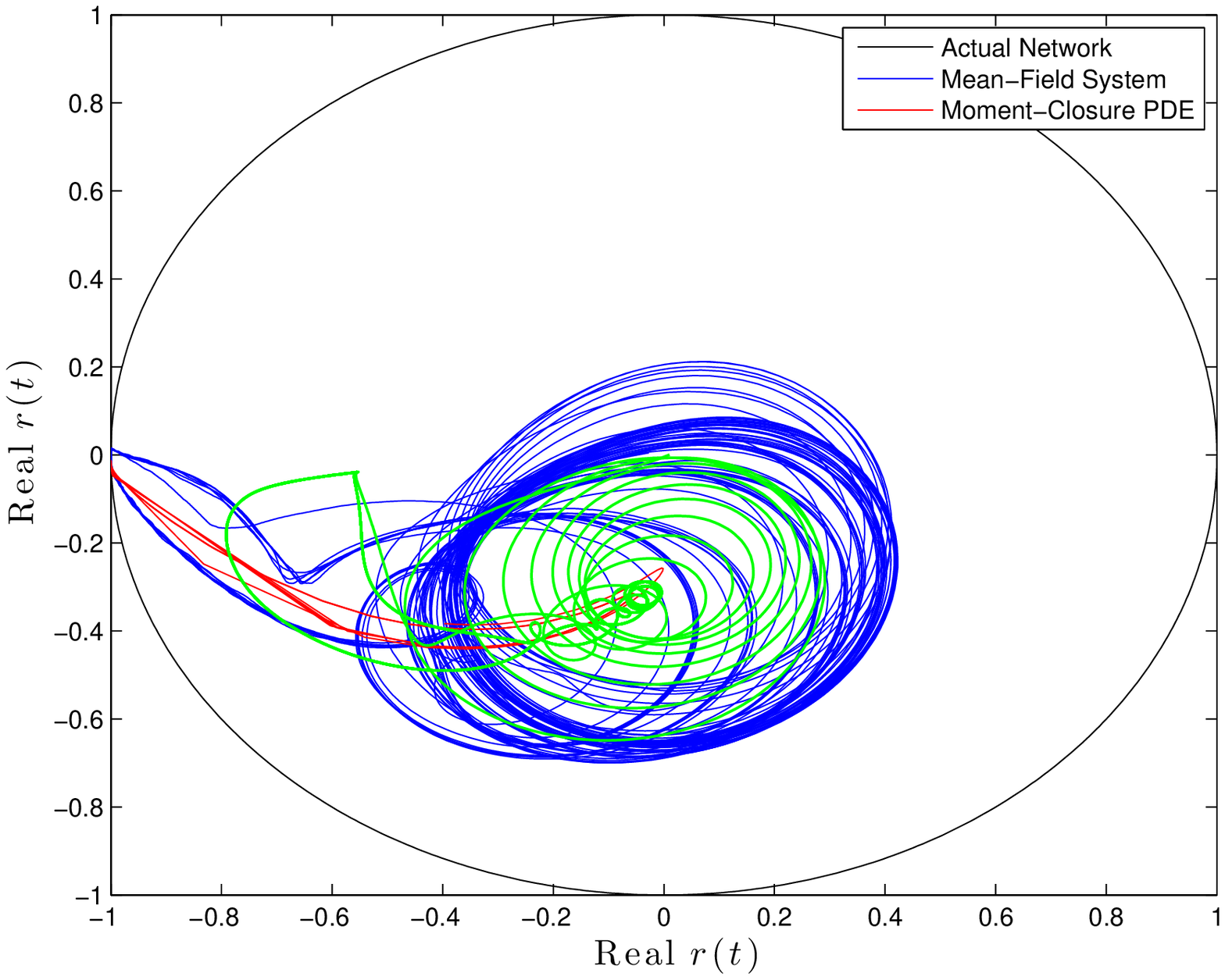}
                \caption{Order Parameter, CH}
                \label{CH2}
        \end{subfigure}
\\

\caption{Comparison of direct numerical simulations of large coupled networks of Izhikevich neurons with noise, the mean field system and the moment closure 
PDE system.  The direct simulations are shown in blue, while the mean-field system is shown in red, and the first order moment closure PDE is shown in green. (a),(c) Network mean variables; (b),(d) order parameter as defined in eq. 
\eqref{orderparam}.  The PDE system has substantially less frequency error than the mean-field system and gives a better representation of the amount of synchronization in the network.  The parameter sets are those of an  intrinsically bursting neuron (a),(b) and a chattering neuron (c),(d).  The values can be found in table \ref{table_param}.  
The standard deviation for the noise is $\sigma = 0.02$ for the intrinsically bursting network, and $\sigma = 0.014$ for the chattering neuron network with the other parameters being $g=0.33, I=0.037$ and $g=0.56$ and $I=0.055$, respectively.  
} \label{fig1}
\end{figure}

\begin{figure}
        \centering
        \begin{subfigure}[b]{0.47\textwidth}
                \centering
                \includegraphics[width=\textwidth]{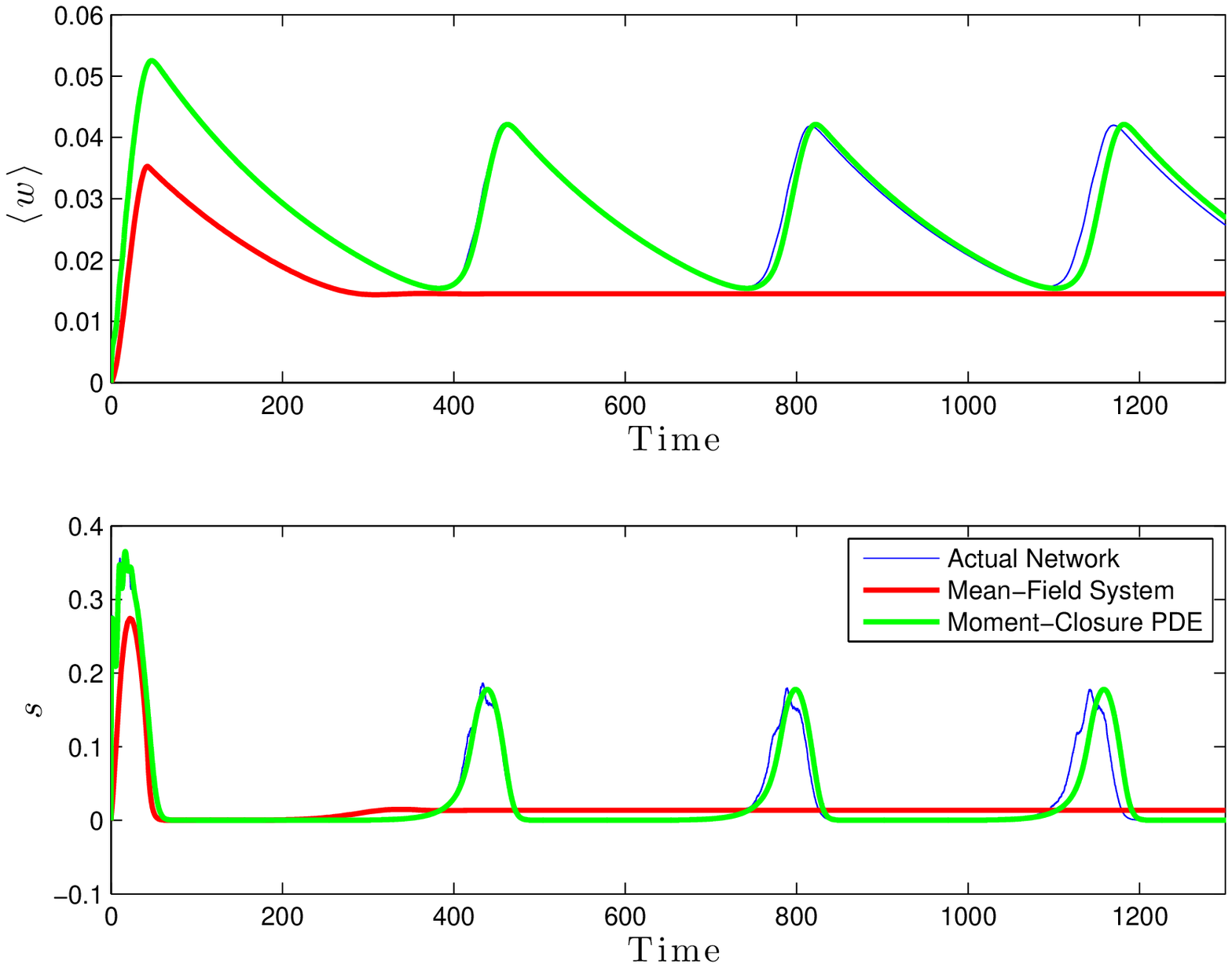}
                \caption{Izhikevich Model with $k$-switching}
                \label{KS1}
        \end{subfigure}
\quad 
        \begin{subfigure}[b]{0.47\textwidth}
                \centering
                \includegraphics[width=\textwidth]{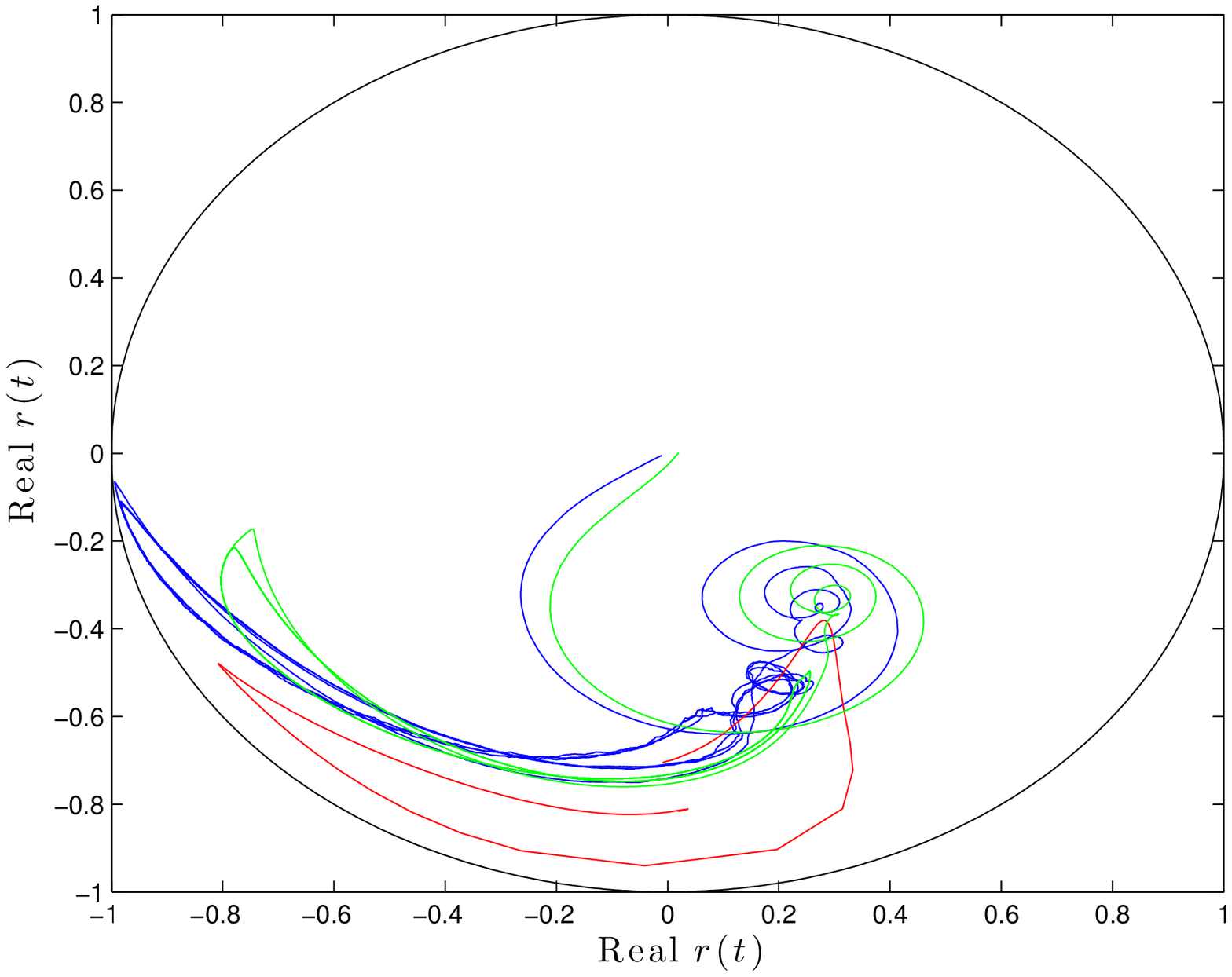}
                \caption{Order Parameter, $k$-switching}
                \label{KS2}
        \end{subfigure}
\\
        \begin{subfigure}[b]{0.47\textwidth}
                \centering
                \includegraphics[width=\textwidth]{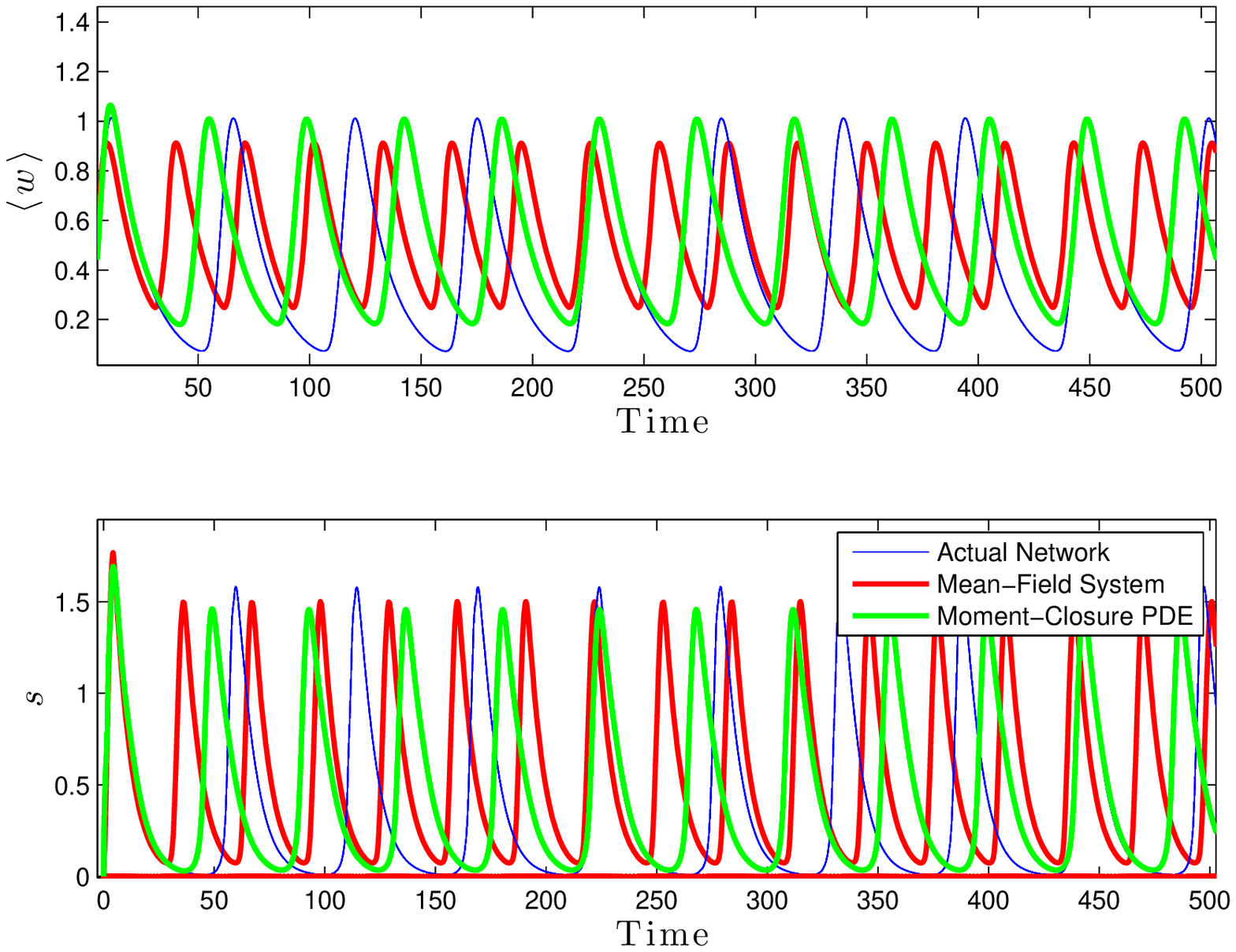}
                \caption{Izhikevich Model, FS Parameters}
                \label{FS1}
        \end{subfigure}
\quad 
        \begin{subfigure}[b]{0.47\textwidth}
                \centering
                \includegraphics[width=\textwidth]{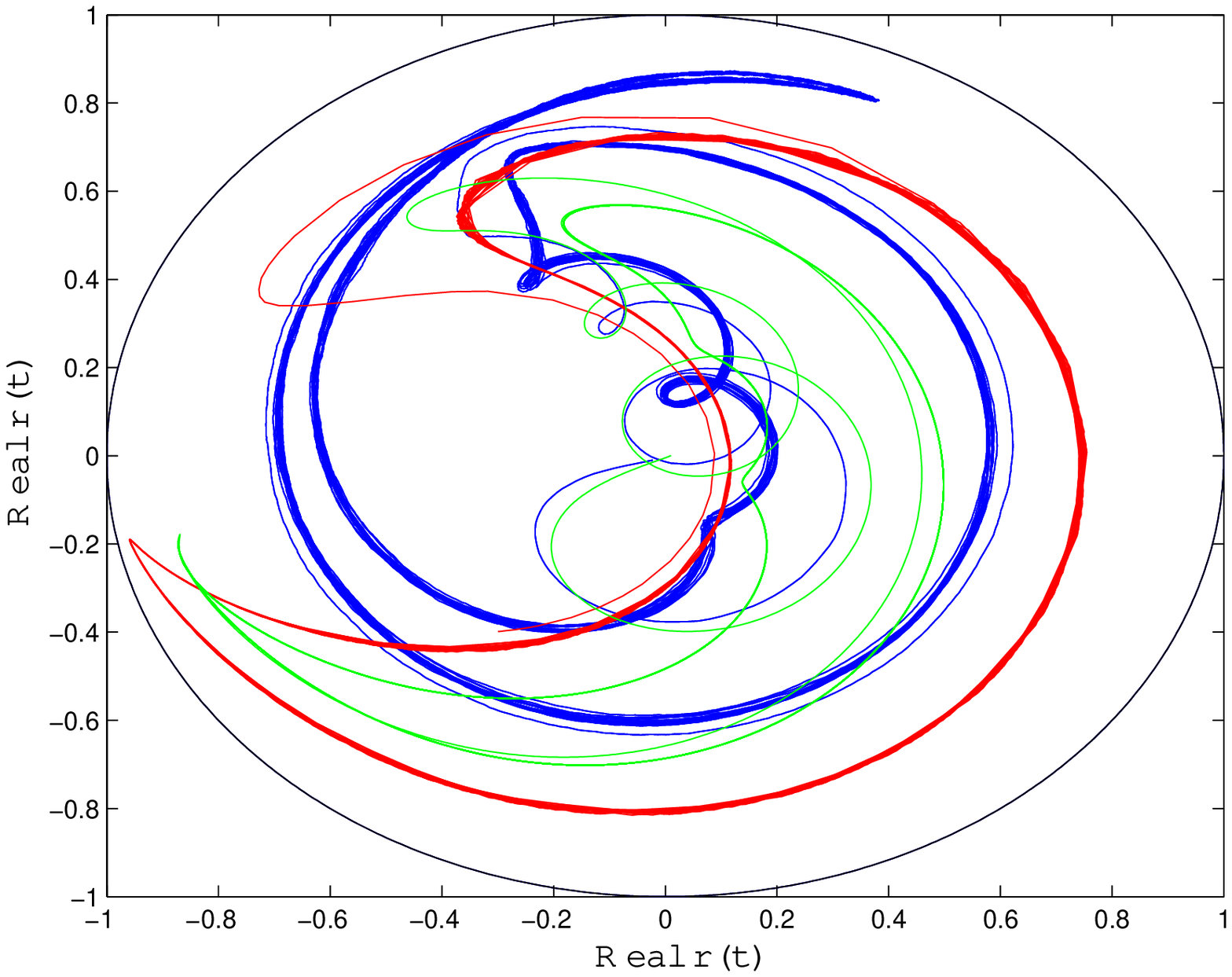}
                \caption{Order Parameter, FS}
                \label{FS2}
        \end{subfigure}

\caption{Comparison of direct simulations of large coupled networks of networks of Izhikevich neurons with noise, the mean field system and the moment closure PDE. (a) the model with $k$-switching, defined by eq. \eqref{kswitching}, to accurately represent spike half-widths.  (c) the model for fast spiking interneurons which has nonlinear $w$ dynamics given by eq. \eqref{nlweq}.   The standard deviation of the noise noise is $\sigma =0.1$ for the fast spiking network, with $g=1.81$ and $I=0.0661$ with the parameter $v_b =0$.  For the $k$-switching network, the parameter values used were $\sigma = 0.032$, $I=0.0189$, $g=0.7692$ in addition to $k_{min} = 0.03$.  The other parameters can be found in Table \ref{table_param}  The direct simulations are shown in blue, while the mean-field system is shown in red, and the first order moment closure PDE is shown in green.  As with the plain Izhikevich model, the PDE has substantially less frequency error than the mean-field system.  The order parameter for the networks, as defined by \eqref{orderparam}, is shown in (b), (d). While not perfect, the moment-closure reduced PDE provides substantially more information about network synchrony than the mean-field system.
}\label{fig2}
\end{figure}

\begin{figure}
        \centering
        \begin{subfigure}[b]{0.6\textwidth}
                \centering
                \includegraphics[width=\textwidth]{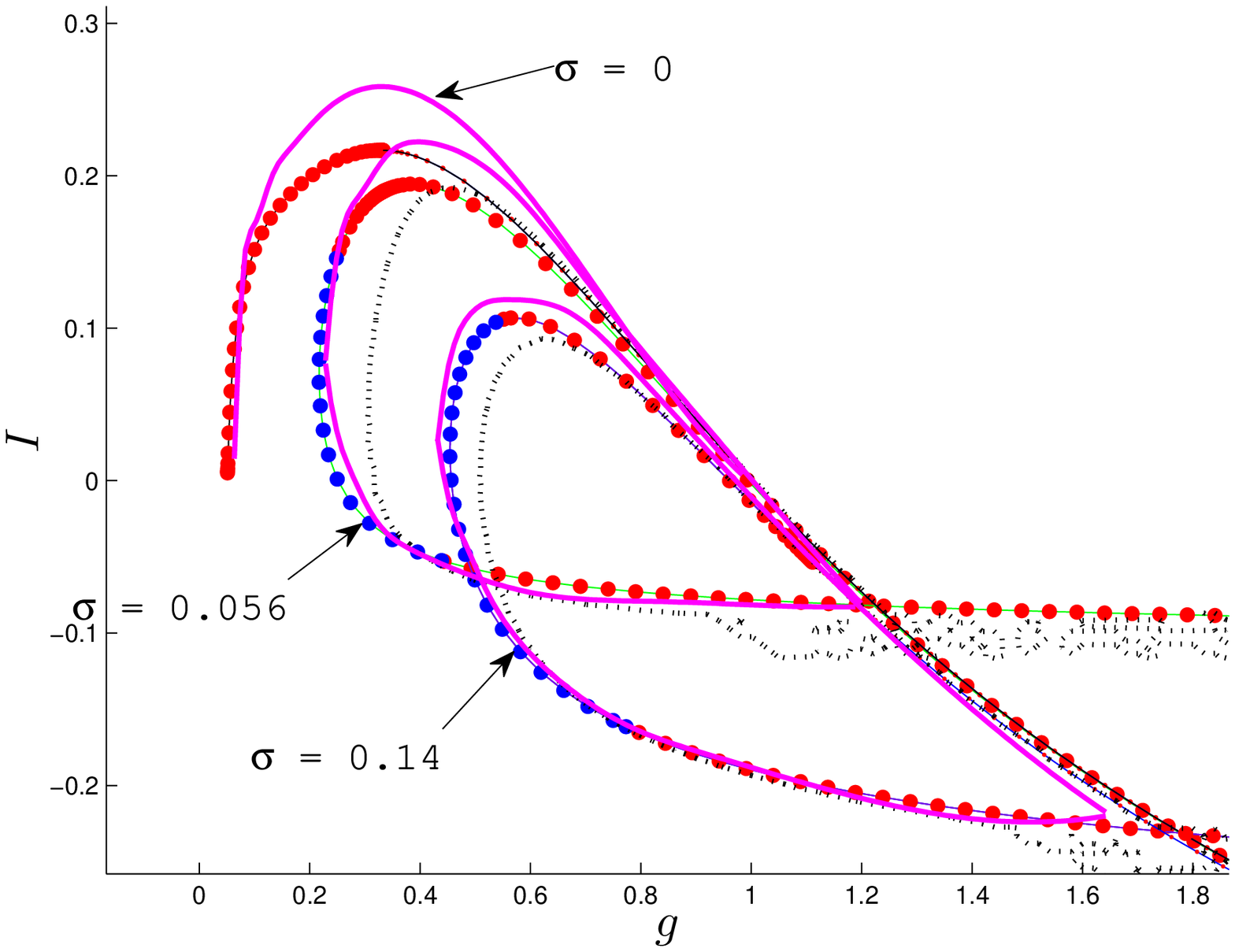}
                \caption{CH parameter set}
                \label{CHbif}
        \end{subfigure}
\\
        \begin{subfigure}[b]{0.6\textwidth}
                \centering
                \includegraphics[width=\textwidth]{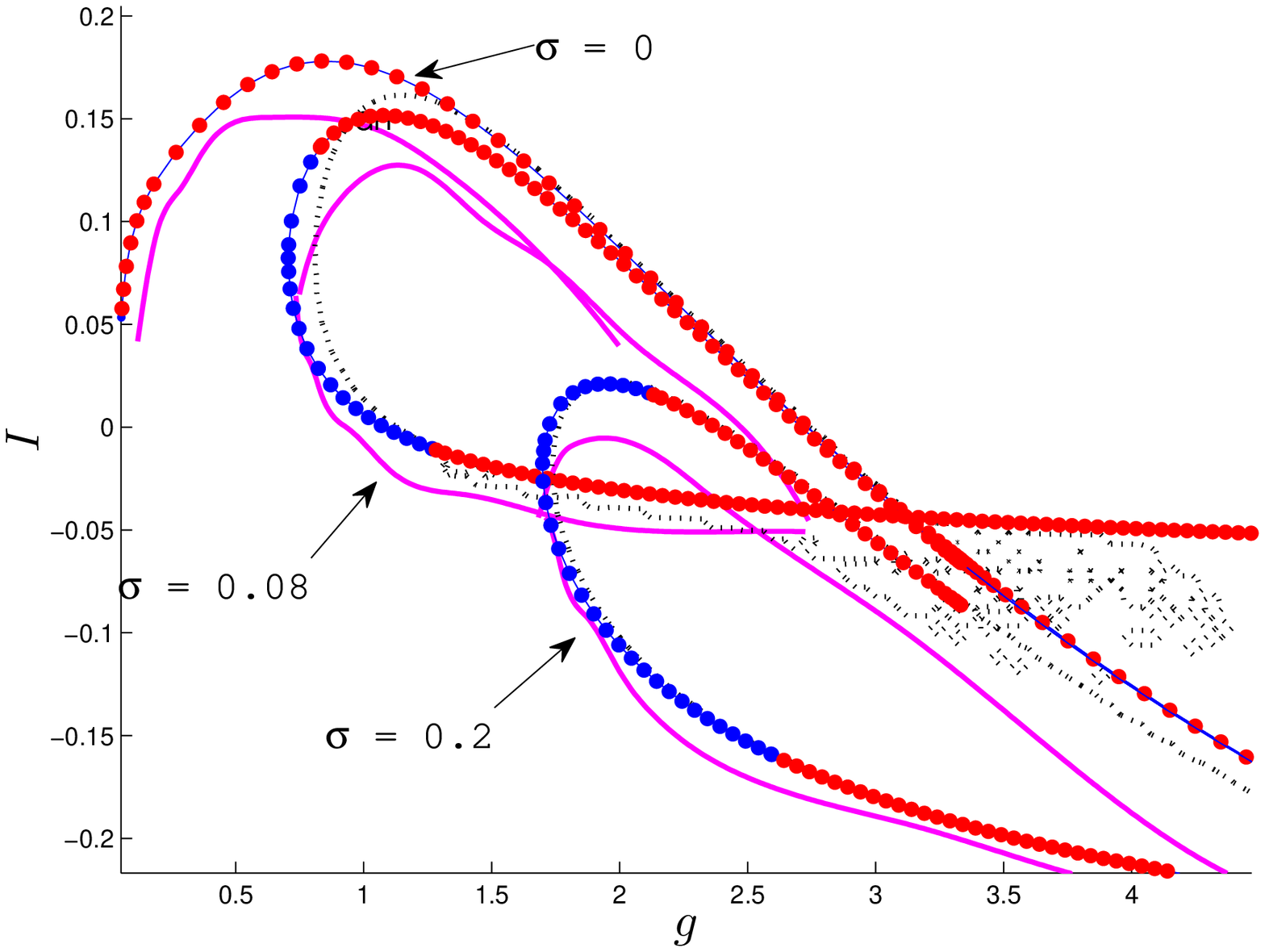}
                \caption{IB parameter set}
                \label{IBbif}
        \end{subfigure}

\quad

\caption{Two parameter bifurcation diagram, generated with various methods, 
for a noisy network of Izhikevich neurons with the parameter sets of 
(a) the chattering neuron and (b) the intrinsically bursting neuron.  
(See columns CH and IB in Table \ref{table_param} for parameter values.)
Magenta curves: A network of 2000 neurons was simulated over a two parameter mesh in the $(g,I)$ parameter space for various $\sigma$ (see labels).  The network was classified as bursting or non-bursting by using a peak finding algorithm on $s(t)$ and $\langle w \rangle$, computed by averaging $w_i(t)$ over the network.  A spline boundary was then fit to the bursting region manually.  
Thick coloured dots: MATCONT was used to numerically continue the Hopf bifurcation curve (see Appendix B).  There are two Bautin bifurcation points for networks with noise that separate the two-parameter Hopf curve into subcritical (red dots) and supercritical (blue dots) branches.  The noiseless network only contains a subcritical branch of Hopf bifurcations.  Black dotted lines: The Real $\lambda  = 0$ contour generated by eqs. (\ref{sc1})-(\ref{scend}) and eqs. (\ref{linstab_syst})--(\ref{stab_matrix}) was used to estimate the 
bursting region. The $\sigma=0$ case was omitted for reasons outlined in the text. }\label{biffig}
 
\end{figure}

  If we further assume that $\langle w | v\rangle$ is a function that is bounded within the interval $[\langle w|v_{peak}\rangle,\langle w | v_{reset}\rangle]$ then it follows that
 \begin{equation}
\langle w | v_{peak} \rangle \leq  \langle w | v\rangle   < \langle w | v_{reset}\rangle = \langle w | v_{peak} \rangle+ w_{jump}.
\end{equation}
In this case the conditional moment $\langle w|v\rangle$ is contained within an interval of size $[w_{jump}]$ for $ v_{reset} \leq v\leq v_{peak}$ and thus after multiplying by $\rho_V(v,t)$ and integrating with respect to $v$, 
we have: 
\begin{equation}
\langle w|v_{peak}\rangle \leq \langle w \rangle \leq \langle w | v_{peak}\rangle + w_{jump}.
\end{equation} 
Thus, the approximation $\langle w | v\rangle = \langle w \rangle$ is valid so long as $\langle w | v\rangle$ is bounded to the $w_{jump}$ interval and $w_{jump}$ is small, as both the conditioned and unconditioned moments lie in the same $w_{jump}$ sized interval.   This is shown in Figure \ref{FIGWV}. It is most apparent in the tonic 
firing regime shown in Figure \ref{figwv1}.

When many of the neurons are quiescent, the boundary condition 
for the first order moment closure system analogous to (\ref{bc1}) 
is difficult to observe numerically as, for a finite network with noise, very few of the neurons are firing.  We have found numerically, however, that in these regimes, the bounds on $\langle w | v\rangle$ still hold as $\langle w | v_{reset}\rangle<\langle w | v_{peak}\rangle + w_{jump}$, and in fact the interval boundary on the conditional moment is significantly smaller than $w_{jump}$.  This is due to the fact that when $I<I^*(s,\langle w \rangle)$, many of the neurons are synchronized around the same stable pseudo-equilibrium in the voltage.
This is shown for example in Figure \ref{figwv4} where the bulk of the neurons are contained in the interval $[0.33,0.365]$ at $t=200$.    Effectively, the neurons are synchronized around $v_-(s,\langle w \rangle)$ aside from a small amount of noise induced firing.  In this situation, first order moment closure contributes even less to the error in these regions.  

\bibliography{noise-NLC}

\section{Figures}

\end{document}